\documentclass{aastex61}

\usepackage{epsfig,graphicx,natbib,url,epstopdf,longtable,amsmath}

\begin{document}

\title{Evolution of cometary dust particles to the orbit of the Earth: particle size, shape, and mutual collisions}

\author{Hongu Yang}
\affiliation{Department of Physics and Astronomy, Seoul National University, 599 Gwanak-ro, Gwanak-gu, Seoul 151-742, Republic of Korea}
\affiliation{Korea Astronomy and Space Science Institute (KASI), Republic of Korea}
\email{hongu@astro.snu.ac.kr}

\author{Masateru Ishiguro}
\affiliation{Department of Physics and Astronomy, Seoul National University, 599 Gwanak-ro, Gwanak-gu, Seoul 151-742, Republic of Korea}
\email{ishiguro@astro.snu.ac.kr}

\correspondingauthor{Masateru Ishiguro}

\begin{abstract}

In this study, we numerically investigated the orbital evolution of cometary dust particles, with special consideration of the initial size frequency distribution (SFD) and different evolutionary tracks according to initial orbit and particle shape. We found that close encounters with planets (mostly Jupiter) are the dominating factor determining the orbital evolution of dust particles. Therefore, the lifetimes of cometary dust particles ($\sim$250 thousand years) are shorter than the Poynting-Robertson lifetime, and only a small fraction of large cometary dust particles can be transferred into orbits with small values of $a$. The exceptions are dust particles from 2P/Encke and, potentially, active asteroids that have little interaction with Jupiter. We also found that the effect of dust shape, mass density, and SFD were not critical in the total mass supply rate to the Interplanetary Dust Particle (IDP) cloud complex when these quantities are confined by observations of zodiacal light brightness and SFD around the Earth's orbit. When we incorporate a population of fluffy aggregates discovered in the Earth's stratosphere and the coma of 67P/Churyumov-Gerasimenko within the initial ejection, the initial SFD measured at the comae of comets (67P and 81P/Wild 2) can produce the observed SFD around the Earth's orbit. Considering the above effects, we derived the probability of mutual collisions among dust particles within the IDP cloud for the first time in a direct manner via numerical simulation and concluded that mutual collisions are mostly ignorable.

\end{abstract}


\section{Introduction}   \label{sec:introduction}

During the first half of 20th century, it became evident that dust particles ejected from solar system objects that are not the Sun itself are scattered in interplanetary space and are therefore observed as zodiacal light (by reflecting sunlight) and meteors (through encounters with the Earth) \citep[and references therein]{2001InterpD..Fech}. At the same time, it was found out that because of the Poynting-Robertson (P--R) effect and solar wind drag, those particles are falling to the Sun on a timescale shorter than $\sim 10^{7}$ years, which is $\approx$2--3 orders of magnitude shorter than the age of the solar system \citep{2006A&ARv..13..159M}. Therefore, recent or continuous dust sources are required to explain the existence of the Interplanetary Dust Particle (IDP) cloud complex at the current epoch. For the source of the IDPs, \citet{2010ApJ...713..816N} argued that approximately 90 \% of the IDPs are originated from comets by connecting the vertical brightness profiles of observed mid-infrared zodiacal light with those of their numerical models. \citet{2015ApJ...813...87Y} came to a similar conclusion (i.e., $>$90 \% from comets) through a comparison between the observed optical properties (i.e., albedo and optical spectral gradient) of zodiacal light and those of different kinds of minor solar system bodies. In contrast, there are studies that have suggested a non-negligible fraction of asteroidal particles in the IDP cloud. \citet{2008Icar..194..769I} examined the Doppler shifts of zodiacal light's Mg I Fraunhofer line and contrived an IDP cloud model that consisted of $30 - 50$\% of asteroidal particles. \citet{2017PASJ...69...31K} analyzed the UV--optical spectrum of zodiacal light taken with the Hubble Space Telescope and insisted that the spectrum is similar to that of C-type asteroids.

Under these circumstances, \citet{2011ApJ...743..129N} reproduced observed helion meteor orbital distribution from dust ejected from Jupiter Family Comets (JFCs). However, as \citet{2017AJ....153..232U} indicated, recent in situ measurements by \textit{Rosetta} at the coma of 67P/Churyumov-Gerasimenkoby \citep{2015Sci...347a3905R,2015ApJ...802L..12F,2016ApJ...816L..32H,2016ApJ...821...19F,2016Natur.537...73B,2016MNRAS.462S..78A,2016MNRAS.462S.304M,2016A&A...596A..87M} suggested that initial dust density and size-frequency distribution (SFD) are different from those in \citet{2011ApJ...743..129N}'s initial condition. Furthermore, the in situ measurements by \textit{Giotto} mission \citep{1995A&A...304..622F} and \textit{Stardust} mission \citep{2004JGRE..10912S04G,2007ESASP.643...35G} as well as the IR observation \citep{2010AJ....139.1491V} unanimously determined SFD similar to that determined by \textit{Rosetta} mission, even though the authors studied different comets (1P/Halley, 81P/Wild 2, and 73P/Schwassmann-Wachmann 3). Therefore, a careful treatment or explanation of these initial conditions is required to further confirm the cometary origin of IDPs. In this work, we investigated this question by introducing fluffy aggregates discovered in the Earth's stratosphere \citep{2003TrGeo...1..689B} and the coma of 67P/Churyumov-Gerasimenko \citep{2015ApJ...802L..12F,2016Natur.537...73B,2016MNRAS.462S.304M}.

Total mass budget of the IDP cloud complex is also an important factor in characterizing the various physical processes because the budget is determined by a balance of every dust supply and removal processes in the solar system. \citet{2011ApJ...743..129N} estimated the total mass ejection rate to the IDP cloud complex as $10^{3} - 10^{4}$ kg s$^{-1}$. In this work, we revisited this budget with modified initial conditions.

Regarding the sink, it has been regarded that catastrophic mutual collisions would be a dominant mechanism breaking IDPs of the size range of $>$200 $\mu$m \citep{1978codu.book..527D, 1985Icar...62..244G, 1986MNRAS.218..185S}. Whereas these studies were targeted for IDPs on fixed circular orbits, their results were considered in previous dynamical research about IDPs on eccentric cometary orbits \citep{2009Icar..201..295W, 2011ApJ...743..129N, 2014ApJ...789...25P}. However, even though these authors did not pursue the reason, recent studies have reported that their results can be explained better when the actual collisional lifetime is longer than that in \citet{1985Icar...62..244G} \citep[for example]{2011ApJ...743..129N, 2016Icar..266..384J}. Under this circumstance, \citet{2016pimo.conf..284S} calculated collisional lifetime of IDPs on fixed eccentric orbits, and reported collisional lifetime longer than \citet{1985Icar...62..244G}. Therefore, it is time to investigate the probability of mutual collisions under more realistic situations, IDPs on orbits which are initially eccentric and evolving under planetary perturbation and radiational acceleration.

In this study, we numerically examined the evolution of dust particles ejected from cometary nuclei. Through this work, we tested a wide variety of initial conditions, namely orbits, dust shape, density and the SFD. We investigated different evolutionary tracks according to initial orbits and dust particle cross-section to mass ratios. Then, we searched for valid combinations of dust particles with different particle shape, density and initial SFD values, considering the in situ measurements by the spacecrafts. The required mass budgets of the IDP cloud complex were derived for valid cases. Finally, we estimated the probability of mutual collisions among dust particles. Note that this is the first attempt to connect the dust SFD of cometary comae with the SFD around the Earth. This work is also the first attempt to derive the probability of mutual collisions in a direct manner via numerical simulation. The description of our methodology is presented in section 2. In section 3.1, We examine different evolutionary tracks under different initial conditions: particles ejected from JFCs or 2P/Encke--like objects; particles with small or large radiative acceleration to gravitational acceleration ratios. In section 3.2, we show the possible combinations of evolutionary tracks under different initial conditions, including particles that are compact or fluffy, particles with high or low particle densities, and different forms of initial SFD. We show the expected mass supply rate for the possible combinations in section 3.3. We estimate the probability of mutual collisions among IDPs with respect to the source orbits, particle sizes, and particle shape in section 3.4. In section 4, we considered relative contribution from different initial orbits, particle size, and particle shape to the zodiacal light brightness and density near the Earth's orbit along with the total mass supply rate.

\section{Methodology}   \label{sec:Methodology}

\subsection{Numerical Integration}   \label{subsec:Numerical Integration}

In our model, we ejected hypothetical dust particles from the orbits of selected 'actual' comets. We considered nine different $\beta$ values, ratios of the solar radiative acceleration with respect to the solar gravitational acceleration: $\beta$ = 0.57, 0.285, 0.114, 0.057, 0.0114, 0.0057, 0.00285, 0.00114, and 0.00057. When spherical particles with a density of 0.8 g cm$^{-3}$ is assumed, these $\beta$ values are equivalent to diameters of 2.5, 5, 12.5, 25, 125, 250, 500, 1250, and 2500 $\mu$m \citep{1979Icar...40....1B}. The details about source selection and dust ejection will be explained in the next subsections. All dust particles were ejected at the same epoch, a Julian date of 2457054.5 (A.D. 2015, February 1, 0:00) for convenience in our simulation. Planetary ephemerides for each epoch were calculated by an N-body simulation using the initial data in \citet{1999MNRAS.304..793C}. During the numerical calculation of the orbital evolution of dust particles, we accounted for the Sun and eight planets as massive objects that exert gravitational accelerations and ignored the gravitational effects of other objects. In addition, radiative acceleration, including P--R drag due to solar radiation, was considered, while these effects due to other sources were ignored because of their weakness. Solar radiation was treated as constant over time. The effect of solar wind drag is assumed to be proportional to 30 \% of the P--R drag \citep{1979Icar...40....1B}. We employed a numerical integrator applying the RADAU15 algorithm \citep{1985dcto.proc..185E}, originally coded by \citet[MERCURY 6.2]{1999MNRAS.304..793C}, and modified by \citet{2014SNUJeong} for taking into account the radiation effects. Dust particles were excluded from the calculation when the heliocentric distance became larger than 80 au or smaller than 0.05 au. The initial integration time step was 3.6525 days, and the variable time step was chosen to accomplish an accuracy of at least $\mid\Delta E/E\mid \ \leq 10^{-12}$ in energy during a single step. We stopped integrations after 2 million years, when most dust particles were excluded from the numerical integration by the above conditions.

\subsection{Source Population}

According to previous research \citep{2006Icar..182..527T, 2011MNRAS.414..458S}, even though JFCs have not been completely surveyed, it is suggested that the distribution of known JFCs is less observationally--biased, and therefore, the SFDs of all known JFCs and those of JFCs with small perihelion distances are statistically same \citep{2013Icar..226.1138F}, implying that known JFCs can statistically represent the whole population. Therefore, supposing that JFCs are the main source of the IDP cloud complex, we began our simulation from existing cometary nuclei. Furthermore, we assumed that number of comets has been constant over a few million years as we explain in section 2.4. This approach of source selection is basically similar to but slightly different from the idea in \citet{2009Icar..201..295W}. The difference in source selection between \citet{2009Icar..201..295W} and this work lays in three parts. Firstly, they assumed that all asteroid population in their model eject a small amount of dust particles as a result of mutual collisions, whereas we regarded that active asteroids (instead of entire asteroid population) eject dust particles. Secondly, we considered small dust particles with large cross-section (i.e., $\beta$$>$0.057) which were not considered in \citet{2009Icar..201..295W}. Lastly, we included 3 times as many comets as \citet{2009Icar..201..295W} did. We chose the source comets from the JPL Horizons comet list as of 2015 January 23 (http://ssd.jpl.nasa.gov/dat/ELEMENTS.COMET).

Among the 3,321 comets in the list, we chose periodic comets with eccentricity $e < 1$ and excluded fragments to avoid duplication from dust particles from the same comets. In total, we chose 1,049 comets for consideration. We classified these comets into JFCs, Encke Type Comets (ETCs), Chiron Type Comets (CTCs) and Halley Type Comets (HTCs) following the criteria in \citet{1996ASPC..107..173L}. HTCs were further classified into two classes: one with the semimajor axis $a$ larger than that of Jupiter $a_\mathrm{J}$ (HTC-1) and another with $a\leq a_\mathrm{J}$ (HTC-2). ETCs were further categorized into two classes: one case having a Tisserand parameter with respect to Jupiter of $T_\mathrm{J}<3.01$ (ETC-1), another with $T_\mathrm{J}\geq3.01$ (ETC-2). In our simulation, dust particles from ETC-1 experienced close encounters with Jupiter and had similar evolutionary tracks to dust particles from JFCs. Dust particles from ETC-2 are less influenced by close encounters with Jupiter. Note that this criteria is different from the working definition of active asteroids (AA), that is, $T_{J}\geq$ 3.08 \citep{2015aste.book..221J} because our concern is not the orbital similarity with asteroids but the dynamical interaction with the Jupiter. The orbit of 2P/Encke is largely different from AAs (occasionally referred to as main-belt comets) in that 2P/Encke has a large eccentricity and a short perihelion distance. However, dust particles from 2P/Encke are less susceptible to close encounter with the Jupiter, similar to AAs. For this reason, both 2P/Encke and AAs are categorized into the same group (ETC-2) in our definition.

In orbital elements space, we constructed 17 step bins for semimajor axis $a$ (0.0--0.5, 0.5--1.0, 1.0--1.5, 1.5--2.0, 2.0--2.5, 2.5--3.0, 3.0--3.5, 3.5--4.0, 4.0--4.5, 4.5--5.0, 5.0--7.0, 7.0--9.0, 9.0--14.0, 14.0--19.0, 19.0--24.0, 24.0--30.0, 30.0--40.0 in au), 5 bins for eccentricity $e$ (0.0--0.2, 0.2--0.4, 0.4--0.6, 0.6--0.8, 0.8--1.0), and 6 bins for inclination $i$ (0\fdg0--30\fdg0, 30\fdg0--60\fdg0, 60\fdg0--90\fdg0, 90\fdg0--120\fdg0, 120\fdg0--150\fdg0, 150\fdg0--180\fdg0), and counted the number of cometary orbits within each bin three dimensionally. Under the choice of bins, each box includes 0 to 63 cometary orbits. In each box, we chose comet samples as representatives of whole population of the given orbital class and performed the numerical calculation with their orbital elements. To eliminate sampling bias and reduce computational load, we chose multiple samples in a box if the number of orbits in a box is larger than 5.5\% of the total number of JFCs, ETC-1s, ETC-2s, and HTC-2s. In the assumption, we implied that dust ejected from other comets in the box should experience orbital evolution similar to those of representative comets. Table 1 shows the list of selected comets included in the calculation and their weighting factor $w_{i}$, which is the number of comets whose dust particles are assumed to have similar evolution with the particles from the listed comets. Later in section 2.4, when we made clones of cometary dust particles, the number of clones was determined to be proportional to $w_{i}$. The orbital elements of selected comets, 64 JFCs, 45 HTC-1s, 2 HTC-2s, 3 ETC-1s, 6 ETC-2s, and 6 CTCs are listed in Table 1 along with their $w_{i}$ values.

As mentioned above, we assumed that our sample of JFCs, HTC-2s, ETC-1s, and ETC-2s is free from discovery bias and used the data in making the IDP cloud complex model in section 2.4. CTCs and HTC-1s were not included in the model but are referred to for comparison. The orbital distribution of these comets is shown in Figure 1. As a comparison (see Table 2--3), we also performed a numerical simulation of asteroidal dust particles. We tested a simple situation in which one dust particle per $\beta$ value was ejected from the largest 1933 asteroids ($>$15 km in diameter), considering that these large asteroids are completely detected.

\startlongtable
\begin{deluxetable*}{cccccc}
\tablecaption{List of comets included in our numerical simulation}
\tabletypesize{\scriptsize}
\tablehead{\colhead{Name} & \colhead{Orbital type$^a$} & \colhead{$q^b$ [au]} & \colhead{$e^c$} & \colhead{$i^d$ [$^{\circ}$]} & \colhead{$w_{i}^e$}}
\startdata
3D/Biela & JFC & 0.879 & 0.751 & 13.2 & 6 \\
4P/Faye & JFC & 1.66 & 0.569 & 9.05 & 21 \\
5D/Brorsen & JFC & 0.590 & 0.810 & 29.4 & 3 \\
6P/d'Arrest & JFC & 1.36 & 0.611 & 19.5 & 6 \\
7P/Pons-Winnecke & JFC & 1.24 & 0.638 & 22.3 & 11 \\
9P/Tempel 1 & JFC & 1.53 & 0.512 & 10.5 & 16.67 \\
10P/Tempel 2 & JFC & 1.42 & 0.537 & 12.0 & 16.67 \\
14P/Wolf & JFC & 2.74 & 0.356 & 27.9 & 28 \\
15P/Finlay & JFC & 0.976 & 0.720 & 6.80 & 11 \\
16P/Brooks 2 & JFC & 1.47 & 0.563 & 4.25 & 16.67 \\
17P/Holmes & JFC & 	2.06 & 0.432 & 19.1 & 21 \\
18D/Perrine-Mrkos & JFC & 1.27 & 0.643 & 17.8 & 6 \\
19P/Borrelly & JFC & 1.35 & 0.626 & 30.4 & 2 \\
29P/Schwassmann-Wachmann 1 & JFC & 5.76 & 0.0419 & 9.38 & 5 \\
30P/Reinmuth 1 & JFC & 1.88 & 0.501 & 8.12 & 21 \\
31P/Schwassmann-Wachmann 2 & JFC & 3.42 & 0.194 & 4.54 & 5 \\
32P/Comas Sola & JFC & 2.00 & 0.556 & 9.97 & 4.67 \\
33P/Daniel & JFC & 2.17 & 0.462 & 22.4 & 15 \\
34D/Gale & JFC & 1.18& 0.761 & 11.7 & 9 \\
42P/Neujmin 3 & JFC & 2.01 & 0.585 & 3.99 & 4.67 \\
53P/Van Biesbroeck & JFC & 2.43 & 0.551 & 6.61 & 15 \\
56P/Slaughter-Burnham & JFC & 2.53 & 0.504 & 8.16 & 15 \\
59P/Kearns-Kwee & JFC & 2.36 & 0.475 & 9.34 & 15 \\
63P/Wild 1 & JFC & 1.95 & 0.651 & 19.8 & 21.5 \\
64P/Swift-Gehrels & JFC & 1.38 & 0.690 & 8.95 & 9 \\
65P/Gunn & JFC & 2.87 & 0.261 & 9.24 & 14 \\
72P/Denning-Fujikawa & JFC & 0.784 & 0.819 & 9.17 & 1 \\
76P/West-Kohoutek-Ikemura & JFC & 1.60 & 0.539 & 30.5 & 2 \\
79P/du Toit-Hartley & JFC & 1.12 & 0.619 & 3.15 & 10 \\
91P/Russell 3 & JFC & 2.62 & 0.329 & 14.1 & 14 \\
99P/Kowal 1 & JFC & 4.74 & 0.229 & 4.33 & 17 \\
121P/Shoemaker-Holt 2 & JFC & 3.75 & 0.185 & 20.2 & 4 \\
127P/Holt-Olmstead & JFC & 2.19 & 0.363 & 14.3 & 2 \\
139P/Vaisala-Oterma & JFC & 3.40 & 0.247 & 2.33 & 9 \\
140P/Bowell-Skiff & JFC & 1.97 & 0.692 & 3.84 & 21.5 \\
142P/Ge-Wang & JFC & 2.49 & 0.498 & 12.3 & 4.67 \\
158P/Kowal-LINEAR & JFC & 4.59 & 0.0287 & 7.91 & 4\\
189P/NEAT & JFC & 1.18 & 0.597 & 20.4 & 2 \\
195P/Hill & JFC & 4.44 & 0.315 & 36.4 & 2 \\
206P/Barnard-Boattini & JFC & 0.979 & 0.689 & 32.0 & 1 \\
226P/Pigott-LINEAR-Kowalski & JFC & 1.92 & 0.480 & 46.3 & 3 \\
249P/LINEAR & JFC & 0.511 & 0.816 & 8.43 & 4 \\
254P/McNaught & JFC & 3.21 & 0.312 & 32.6 & 1 \\
269P/Jedicke & JFC & 4.07 & 0.435 & 6.61 & 1 \\
318P/McNaught-Hartley & JFC & 2.48 & 0.671 & 17.6 & 10 \\
C/1999 XS87 (LINEAR) & JFC & 2.77 & 0.841 & 14.8 & 3 \\
C/2001 M10 (NEAT) & JFC & 5.30 & 0.801 & 28.1 & 2 \\ 
C/2002 A1 (LINEAR) & JFC & 4.71 & 0.725 & 14.1 & 4 \\
P/2002 T5 (LINEAR) & JFC & 3.93 & 0.437 & 30.9 & 3 \\
C/2003 E1 (NEAT) & JFC & 3.24 & 0.764 & 33.5 & 1 \\
P/2004 A1 (LONEOS) & JFC & 5.46 & 0.308 & 10.6 & 7 \\
P/2004 V3 (Siding Spring) & JFC & 3.94 & 0.446 & 50.5 & 2 \\
C/2007 S2 (Lemmon) & JFC & 5.56 & 0.557 & 16.9 & 2 \\
C/2008 E1 (Catalina) & JFC & 4.83 & 0.548 & 35.0 & 1 \\
P/2008 O3 (Boattini) & JFC & 2.50 & 0.695 & 32.3 & 2 \\
P/2010 H2 (Vales) & JFC & 3.11 & 0.193 & 14.3 & 1 \\
P/2010 H5 (Scotti) & JFC & 6.03 & 0.156 & 14.1 & 1 \\
C/2011 KP36 (Spacewatch) & JFC & 4.88 & 0.873 & 19.0 & 1 \\
P/2012 C3 (PANSTARRS) & JFC & 3.62 & 0.626 & 9.19 & 9 \\
C/2012 Q1 (Kowalski) & JFC & 9.48 & 0.637 & 45.2 & 1 \\
P/2012 US27 (Siding Spring) & JFC & 1.82 & 0.649 & 39.3 & 1 \\
C/2012 X2 (PANSTARRS) & JFC & 4.75 & 0.771 & 34.1 & 1 \\
P/2013 EW90 (Tenagra) & JFC & 3.30 & 0.196 & 31.8 & 1 \\
2003EH1 & JFC & 1.19 & 0.618 & 70.8 & 1 \\
74P/Smirnova-Chernykh & ETC-1 & 3.54 & 0.149 & 6.65 & 5 \\
87P/Bus & ETC-1 & 2.10 & 0.389 & 2.60 & 8 \\
94P/Russell 4 & ETC-1 & 2.23 & 0.364 & 6.18 & 6 \\
2P/Encke & ETC-2 & 0.336 & 0.848 & 11.8 & 1 \\
233P/La Sagra & ETC-2 & 1.79 & 0.409 & 11.3 & 5 \\
259P/Garradd & ETC-2 & 1.79 & 0.342 & 15.9 & 1 \\
311P/PANSTARRS & ETC-2 & 1.94 & 0.115 & 4.97 & 2 \\
324P/La Sagra & ETC-2 & 2.62 & 0.154 & 21.4 & 1 \\
331P/Gibbs & ETC-2 & 2.88 & 0.0414 & 9.74 & 1 \\
39P/Oterma & CTC & 5.47 & 0.246 & 1.94 & 2 \\
165P/LINEAR & CTC & 6.83 & 0.622 & 15.9 & 1 \\
166P/NEAT & CTC & 8.56 & 0.383 & 15.4 & 2 \\
167P/CINEOS & CTC & 11.8 & 0.270 & 19.1 & 1 \\
P/2005 S2 (Skiff) & CTC & 6.40 & 0.197 & 3.14 & 2 \\
C/2013 C2 (Tenagra) & CTC & 9.13 & 0.429 & 21.3 & 2 \\
1P/Halley & HTC-1 & 0.586 & 0.967 & 162. & 3 \\
8P/Tuttle & HTC-1 & 1.03 & 0.820 & 55.0 & 1 \\
12P/Pons-Brooks & HTC-1 & 0.774 & 0.955 & 74.2 & 2 \\
13P/Olbers & HTC-1 & 1.18 & 0.930 & 44.6 & 4 \\
23P/Brorsen-Metcalf & HTC-1 & 0.479 & 0.972 & 19.3 & 2 \\
27P/Crommelin & HTC-1 & 0.748 & 0.919 & 29.0 & 4 \\
35P/Herschel-Rigollet & HTC-1 & 0.748 & 0.974 & 64.2 & 5 \\
55P/Tempel-Tuttle & HTC-1 & 0.976 & 0.906 & 162. & 2 \\
109P/Swift-Tuttle & HTC-1 & 0.960 & 0.963 & 113. & 1 \\
122P/de Vico & HTC-1 & 0.659 & 0.963 & 85.4 & 2 \\
126P/IRAS & HTC-1 & 1.72 & 0.696 & 45.8 & 1 \\
161P/Hartley-IRAS & HTC-1 & 1.27 & 0.835 & 95.7 & 1 \\
177P/Barnard & HTC-1 & 1.11 & 0.955 & 31.2 & 3 \\
262P/McNaught-Russell & HTC-1 & 1.28 & 0.815 & 29.1 & 1 \\
273P/Pons-Gambart & HTC-1 & 0.810 & 0.975 & 136. & 2 \\
C/1855 L1 (Donati) & HTC-1 & 0.568 & 0.986 & 157. & 1 \\
C/1857 O1 (Peters) & HTC-1 & 0.747 & 0.980 & 32.8 & 3 \\
C/1906 V1 (Thiele) & HTC-1 & 1.21 & 0.949 & 56.0 & 2 \\
C/1921 H1 (Dubiago) & HTC-1 & 1.10 & 0.848 & 22.1 & 1 \\
C/1937 D1 (Wilk) & HTC-1 & 0.619 & 0.981 & 26.0 & 2 \\
C/1998 G1 (LINEAR) & HTC-1 & 2.13 & 0.824 & 110. & 2 \\
C/1998 Y1 (LINEAR) & HTC-1 & 1.75 & 0.924 & 28.1 & 1 \\
C/1999 E1 (Li) & HTC-1 & 3.92 & 0.760 & 46.9 & 1 \\
C/1999 K4 (LINEAR) & HTC-1 & 1.44 & 0.915 & 121. & 1 \\
C/2001 OG108 (LONEOS) & HTC-1 & 0.994 & 0.925 & 80.2 & 1 \\
P/2001 Q6 (NEAT) & HTC-1 & 1.41 & 0.824 & 56.9 & 2 \\
C/2001 W2 (BATTERS) & HTC-1 & 1.05 & 0.941 & 116. & 3 \\
C/2002 B1 (LINEAR) & HTC-1 & 2.27 & 0.771 & 51.0 & 1 \\
C/2002 CE10 (LINEAR) & HTC-1 & 2.05 & 0.791 & 145. & 1 \\
C/2003 F1 (LINEAR) & HTC-1 & 4.01 & 0.806 & 70.2 & 1 \\
C/2003 H2 (LINEAR) & HTC-1 & 2.18 & 0.943 & 74.2 & 1 \\
C/2003 R1 (LINEAR) & HTC-1 & 2.10 & 0.893 & 149. & 2 \\
C/2003 U1 (LINEAR) & HTC-1 & 1.80 & 0.922 & 164. & 1 \\
C/2005 N5 (Catalina) & HTC-1 & 1.63 & 0.943 & 21.4 & 1 \\
P/2006 R1 (Siding Spring) & HTC-1 & 1.67 & 0.702 & 160. & 1 \\
P/2010 D2 (WISE) & HTC-1 & 3.66 & 0.453 & 57.2 & 1 \\
P/2010 JC81 (WISE) & HTC-1 & 1.81 & 0.777 & 38.7 & 1 \\
C/2010 L5 (WISE) & HTC-1 & 0.791 & 0.904 & 147. & 1 \\
C/2011 J3 (LINEAR) & HTC-1 & 1.45 & 0.926 & 115. & 1 \\
C/2011 L1 (McNaught) & HTC-1 & 2.24 & 0.797 & 65.5 & 1 \\
P/2012 NJ (La Sagra) & HTC-1 & 1.29 & 0.848 & 84.4 & 1 \\
P/2013 AL76 (Catalina) & HTC-1 & 2.05 & 0.685 & 145. & 1 \\
C/2013 V3 (Nevski) & HTC-1 & 1.39 & 0.891 & 32.1 & 1 \\
C/2014 W10 (PANSTARRS) & HTC-1 & 7.42 & 0.604 & 73.0 & 1 \\
P/2015 A3 (PANSTARRS) & HTC-1 & 1.15 & 0.848 & 173. & 2 \\
96P/Machholz 1 & HTC-2 & 0.124 & 0.959 & 58.3 & 1 \\
P/1999 J6 (SOHO) & HTC-2 & 0.0491 & 0.984 & 26.6 & 5 \\
\enddata
\tablenotetext{a}{Orbital types of comets (see section 2.2)}
\tablenotetext{b}{Perihelion distance}
\tablenotetext{c}{Eccentricity}
\tablenotetext{d}{Inclination}
\tablenotetext{e}{Initial weighting factor (see section 2.2)}
\end{deluxetable*}

\begin{figure*}
\figurenum{1}
\epsscale{1}
\plottwo{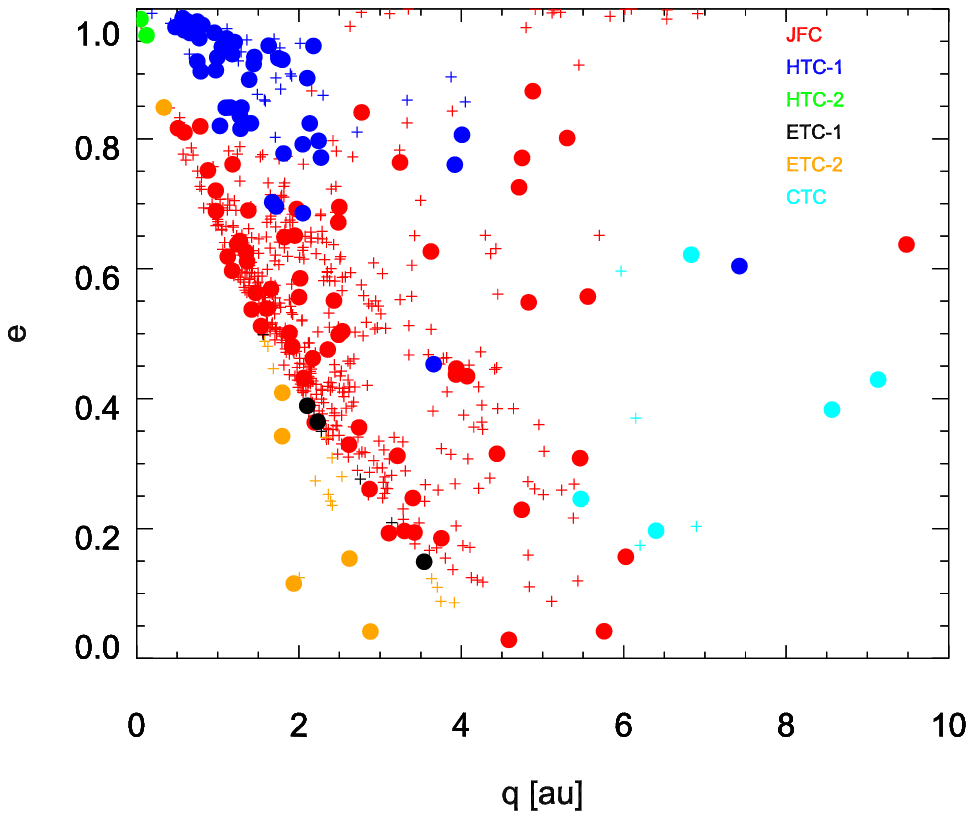}{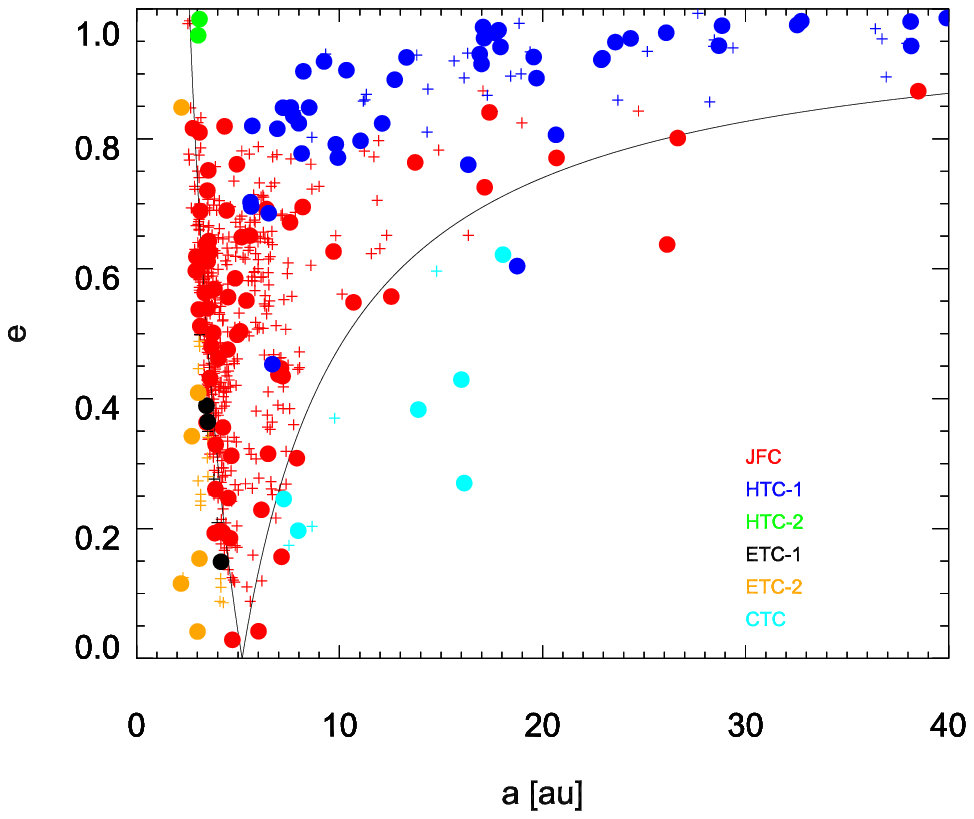}
\plottwo{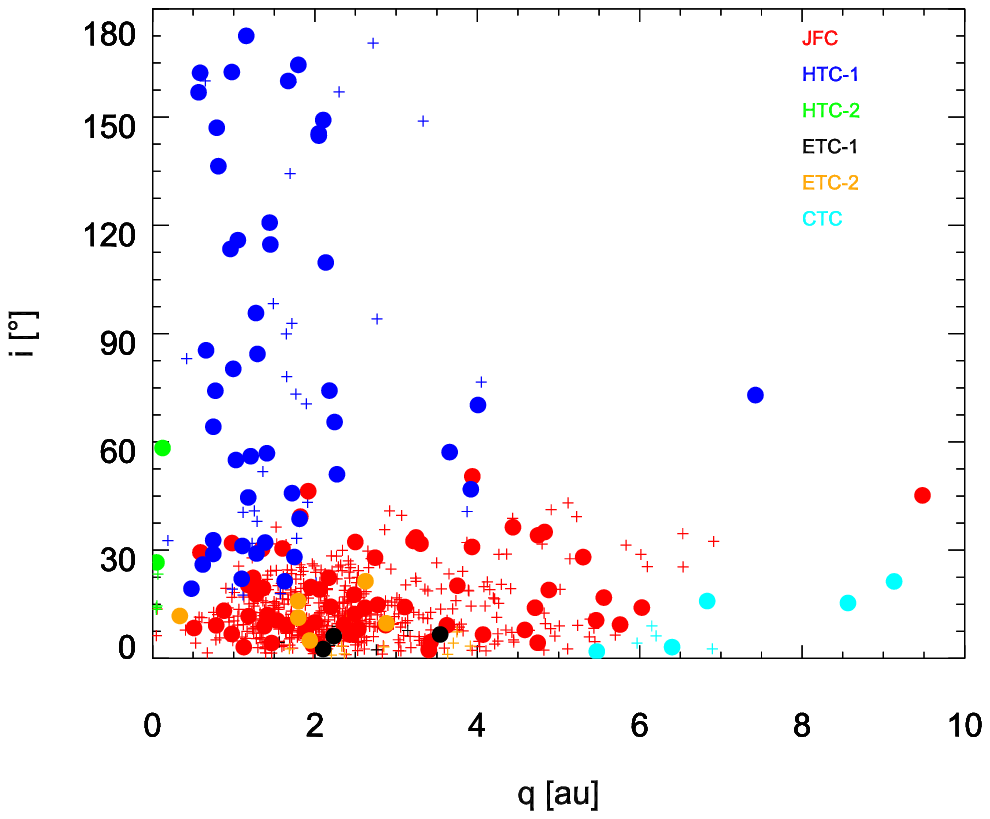}{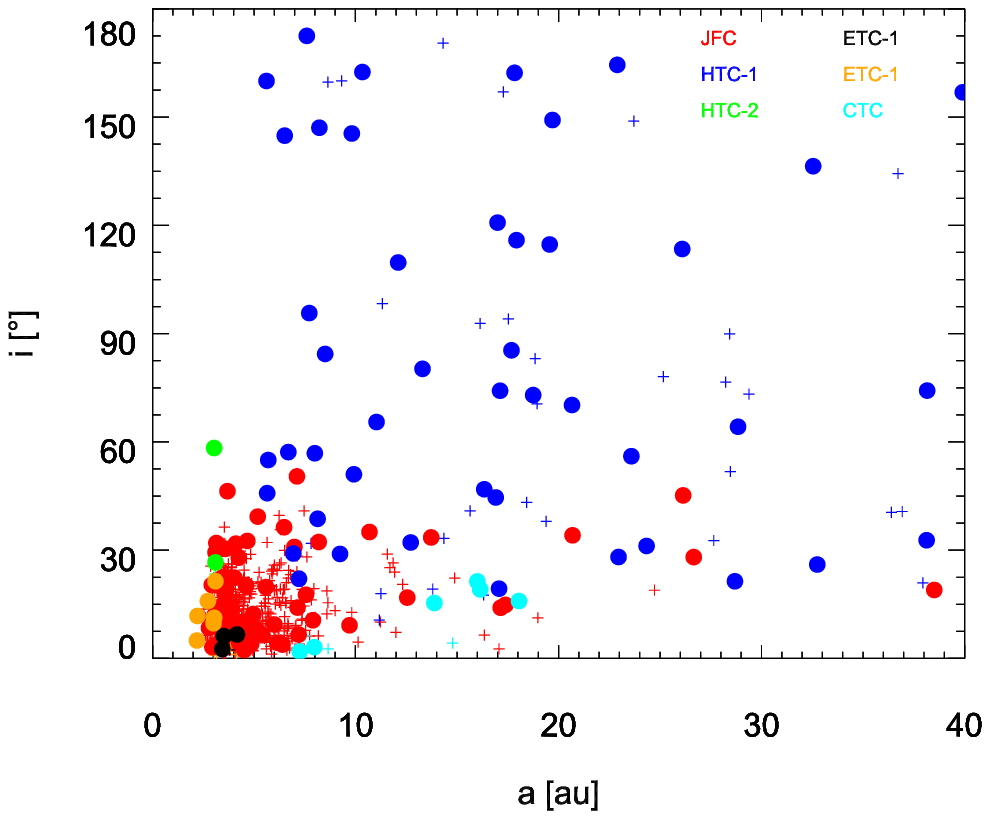}
\caption{Orbital element distributions of all comets with semimajor axes shorter than 40 au, discovered before 2015 January 23. Large filled circles are comets included in our numerical integration, and small crosses are comets that are not included.}
\end{figure*}

\subsection{Dust ejection}

We ejected dust particles from the orbits of the above actual cometary nuclei through the following methods. At first, the orbits of comets at the epoch $t_{0} = $ JD 2457054.5 were numerically calculated from the orbital elements of the JPL Horizons comet lists. Then, five orbital elements of the comets were fixed, except for anomalies. All dust particles were ejected simultaneously at the epoch on the orbits of comets with randomized true anomalies. We ejected 100 dust particles per given size per cometary orbit for particles with $\beta \leq 0.00285$ and 50 for particles with $\beta \geq 0.00114$.

We distributed the true anomalies as the number density of ejected dust particles becomes a function of heliocentric distance following the observed dust ejection efficiency change \citep{2007Icar..189..169I,2009A&A...508.1031M,2012PASJ...64..134H}. Where $N(r_\mathrm{h})$ denotes the initial number density of ejected particles on the given cometary orbit in cm$^{-3}$, and $r_\mathrm{h}$ denotes the heliocentric distance in au, we distributed the particles as:
\begin{equation}
\label{eq:eq1}
N(r_\mathrm{h}) \propto r_\mathrm{h}^{-3.0}~~,
\end{equation}
The ejection velocities $v_{ej}(r_\mathrm{h})$ are given as their directions are randomly distributed over the sunlit hemisphere following an uniform probability distribution, and their speeds are given as a function of $\beta$ and $r_\mathrm{h}$ \citep{2007Icar..189..169I,2012PASJ...64..134H}:
\begin{equation}
\label{eq:eq2}
v_{ej}(r_\mathrm{h}) = 200[\mathrm{m\ s}^{-1}]\beta^{0.5}r_\mathrm{h}^{-0.5}~~,
\end{equation}

Although we assumed a relatively realistic situation in ejecting dust particles compared to the cases with zero ejection velocity, the added ejection velocities numerically function as a kind of cloning process because the ejection velocities are small compared to the orbital velocities. Likewise, even though we ejected dust particles simultaneously at different positions on the cometary orbits instead of the exact position of nuclei at the epoch, dynamically, this approach is the same as changing the anomaly without changing other orbital elements, and therefore, this method works as if it were a cloning process.

The initial orbital elements of ejected dust particles were calculated after the ejection velocities were added to the orbital velocities, taking account of the change of orbital elements due to radiative acceleration.

\subsection{IDP complex modeling}

We assumed that all of the above cometary nuclei eject dust particles at the same rate when we calculate the average over single revolutions. We also assumed that all nuclei eject dust particles with same shape and SFD once we fixed these parameters. Under these assumptions, the ratio between initial dust masses JFCs : HTC-2 : ETC-1 : ETC-2 is 486 : 6 : 19 : 11.

We constructed an IDP distribution covering different ages of particles for given $\beta$ values using the following methods. First, we recorded the positions and velocities of dust particles every 100 years and cloned 1000 $\times$ $w_{i}$ particles from the each individual dust particle. We set the number of the clones to be proportional to the parameter $w_{i}$. During the cloning process, we kept three orbital elements (i.e., the semimajor axes, eccentricities and inclinations) constant and randomized their arguments of perihelion, longitudes of ascending node, and mean anomalies for each clone. Consequently, we made snapshots composed of clones with single $\beta$ values and from the same class of cometary population every 100 years. Assuming a steady state, (i.e., that dust ejections have occurred constantly over two million years), we regarded dust particle records after a given time of integration from the current epoch as records from possible ejections at the same amount of time before the current epoch. This assumption is similar with that made in previous works such as \citet{2009Icar..201..295W}. Finally, the snapshots were summed together, constructing an IDP distribution covering the full dynamical evolution from sources to sinks.

Using the above information from our numerical calculation, we validated the initial dust ejection conditions through following methods. Whenever the initial conditions about dust particle shape, mass density and initial SFD were given, the resulting IDP cloud complex models were derived. We constructed a three--dimensional grid in Cartesian coordinates. 0.1 au $\times$ 0.1 au $\times$ 0.1 au grids were constructed throughout interplanetary space, and the number of dust particles of a given $\beta$ value within a single grid were counted. The results of different $\beta$ values were weighted by the given initial conditions (see section 3.2). As a next step, we derived the cumulative number, cross-section and mass distribution of IDPs in the grid box, assuming a piecewise exponential SFD within every single grid box. Then, after summing the numbers from different source populations, the resulting SFDs around the Earth's orbit were normalized at $\beta = 0.00285$ and compared with the observational SFD in \citet{1985Icar...62..244G}, which was obtained through the \textit{Pegasus} and \textit{HEOS-2} missions and widely applied as the reference SFD model. We examined the ratio between our normalized SFD and the observed one between $\beta$ of 0.00057 $\sim$ 0.114 and treat the given initial conditions as valid if the ratio in SFD is between 0.5 $\sim$ 2 for every $\beta$ value. For such cases, we derived a goodness-of-fit as the averaged absolute values of the logarithm of the ratios.

Finally, the total IDP cloud complex model was scaled to match the observed zodiacal light brightness in the anti-solar direction (i.e., gegenschein) \citep{1998A&AS..127....1L, 2013ApJ...767...75I}. As we mentioned in section 2.1, we limited the largest dust size to $\beta$ = 0.00057. This cut-off is expected to have negligible effects on the total cross-section but may exert a considerable influence on the total mass according to the initial conditions. Therefore, we tabulated the total mass supply in two cases, namely up to $\beta \geq$ 0.00114 and $\beta \geq$ 0.00057, along with validated initial conditions.

\section{Results}   \label{sec:Results}

\subsection{Orbital evolution of dust particles according to different populations and sizes} \label{subsec:Orbital evolution of dust particles according to different populations and sizes}

In this work, the sinks of dust particles were determined by close encounters with planets for the most cases. We will explain the role of close encounter, which concurs with previous research. Dust particles ejected from comets fall to the Sun because of P--R drag if the solar gravity and radiation of the Sun are the only two factors affecting their orbits. P--R drag continuously reduces the $a$ and $e$ of the dust particles. The effect of resonances with planets (i.e., the mean motion and secular resonances) themselves do not change sinks of dust particles significantly because the resonances change $e$ rather than $a$. Note that the lifetime of dust particles may be changed by resonances, which would increase the lifetime of dust particles temporarily by trapping them. In contrast, close encounters with planets, mainly Jupiter, can completely change the orbits of dust particles and usually prevent falls of the dust particles to the Sun. As a consequence of close encounters, $a$ may decrease or increase, and any $e$ value may occur from a circular orbit to a hyperbolic orbit. Dust particles ejected from the JFCs are prone to close encounters with Jupiter because the initial orbits of their source comets intersect or are close to the Jovian orbit and have slow relative velocities to Jupiter.

Smaller dust particles have higher chances of avoiding close encounters than larger particles because they fall to the Sun faster via P--R drag and therefore reside for a shorter period of time in the region where the Jovian gravity is critical. Figure 2 shows the lifetime of dust particles under radiative acceleration. For comparison, we construct the lifetime in the case where there are no planetary perturbations. Because P--R drift is slow for large particles, the lifetime of particles that are larger than 100 $\mu$m is determined by close encounter, not by P--R drift. Therefore, the lifetimes of large dust particles are similar to those of their parent bodies, $\sim$ 100 $-$ 300 thousands years ($\sim$ 100 thousands years in weighted median, $\sim$ 250 thousands years in weighted average) \citep{1994Icar..108...18L, 1997Icar..127...13L}. Note that the lifetime with planetary perturbation included the effects of resonances. The resonances may increase the lifetime of dust particles by temporarily trapping them, especially for the particles which have evolved into the inner solar system.

\begin{figure*}
\figurenum{2}
\epsscale{1}
\plottwo{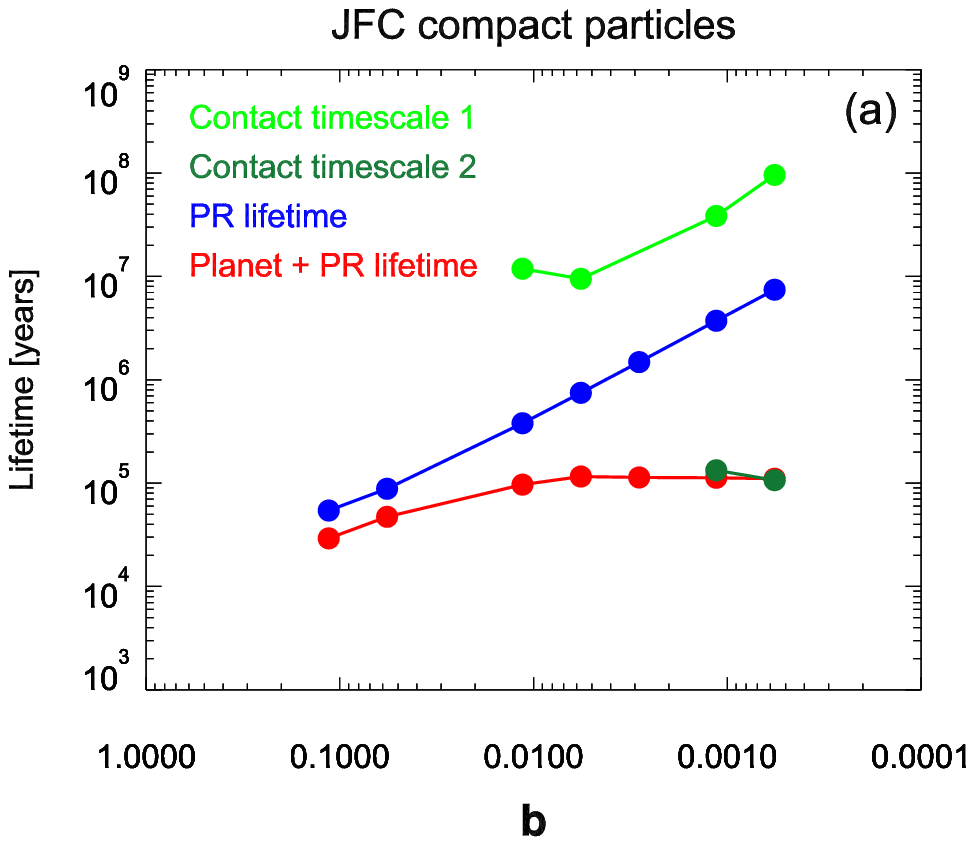}{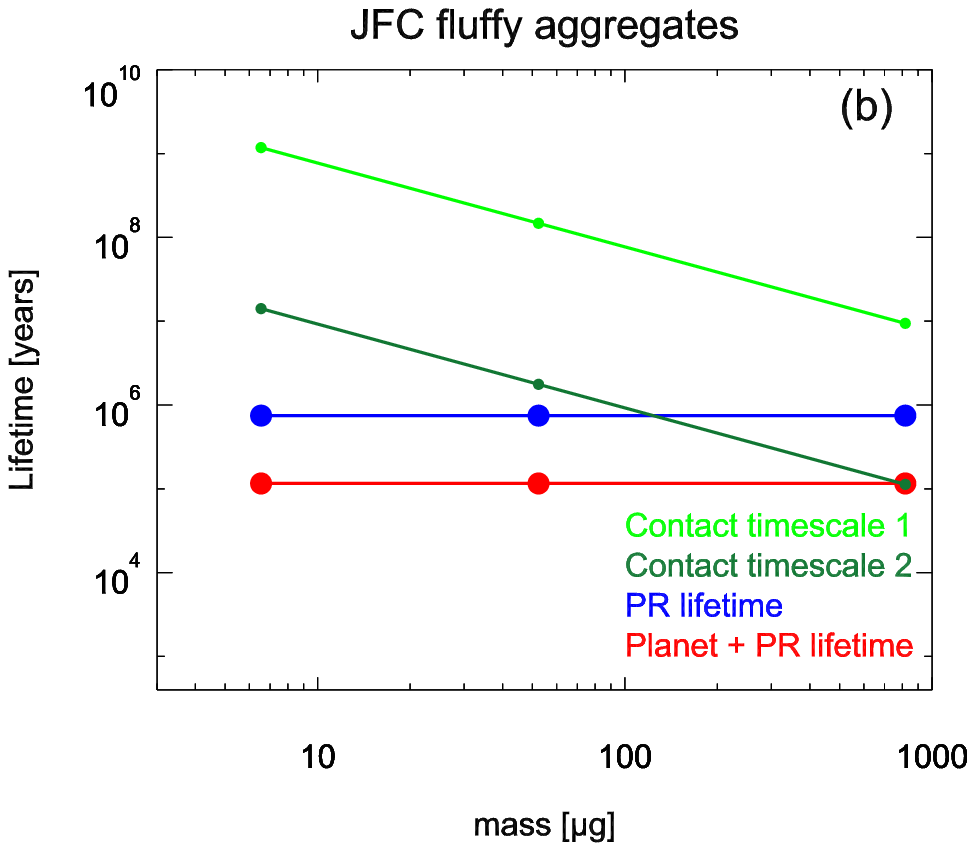}
\caption{Dynamical lifetime, P--R lifetime and contact timescale of cometary dust particles. The values here are the weighted median of all dust particles. Dynamical and P--R lifetimes were calculated by numerical integration of dust particles both with and without planetary gravities. When there are no planets, we integrated the orbits of dust particles until they fall to the Sun without limit. (a) Contact timescales are calculated based on the values in the first line of Table 7. Contact timescale 1 is the timescale required for contact with a projectile 10 times smaller than the particle in diameter, and contact timescale 2 is the value for a projectile 100 times smaller. Impact velocities are not considered. (b) P--R lifetimes of fluffy aggregates are independent from mass, as explained in section 3.2. Contact timescales are calculated using the values in the first line of Table 8. Contact timescale 1 is the timescale required for contact with a projectile with a 125 $\mu$m or larger diameter, and contact timescale 2 is the value for projectile with a 25 $\mu$m or larger diameter.}
\end{figure*}

We measured the fraction of dust particles transferred into the inner solar system (a $\leq$ 1 au) for different populations with different sizes (Table 2). The probability of successful transfer to an orbit with a semimajor axis smaller than 1 au will be selectively higher for dust particles meeting the following two conditions: dust particles ejected from objects the aphelion distances of which are shorter than the semimajor axis of Jupiter and that are free from close encounters with Jupiter, ETC-2s and asteroids; and dust particles that are small. We checked that the dust particles that are not transferred to the inner solar system within 2 million years of integration time satisfy one of the following conditions: trapped in the Trojan region; or $\beta < 0.00114$ and free from planetary encounter and thus still under P--R decay. The latter case includes dust particles ejected from AAs and asteroids, and we performed 5 million years of integration for these particles to complete the dust particle evolution process. The proportions of surviving dust particles after two million years of integration are summarized in Table 3.

\begin{deluxetable*}{ccccccccccc}
\tablecaption{Proportion of dust particles with $a \leq$ 1 AU from different source populations and $\beta$ values [\%].}
\tablehead{\colhead{$\beta$} & \colhead{0.57} & \colhead{0.285} & \colhead{0.114} & \colhead{0.057} & \colhead{0.0114} & \colhead{0.0057} & \colhead{0.00285} & \colhead{0.00114} & \colhead{0.00057}}
\startdata
JFC & 0.025 & 4.3 & 41 & 45 & 30 & 25 & 16 & 5.6 & 2.9 \\
ETC-1 & 0.0 & 9.8 & 47 & 30 & 19 & 12 & 6.9 & 2.6 & 0.84 \\
ETC-2 & 0.0 & 25 & 73 & 77 & 97 & 99 & 100 & 100 & 97 \\
HTC-1 & 0.0 & 0.22 & 3.2 & 13 & 19 & 12 & 5.5 & 1.4 & 0.16 \\
HTC-2 & 0.0 & 0.0 & 0.0 & 0.0 & 0.83 & 1.0 & 0.0 & 0.33 & 0.0 \\
CTC & 0.0 & 0.30 & 7.2 & 12 & 8.1 & 5.2 & 3.1 & 0.80 & 0.0 \\
Asteroids & 0.052 & 35 & 86 & 94 & 97 & 96 & 96 & 87 & 6.5 \\
\enddata
\end{deluxetable*}

\begin{deluxetable*}{cccccccccc}
\tablecaption{Proportion of dust particles that remain in the solar system after 2 million years of integration [\%].}
\tablehead{\colhead{$\beta$} & \colhead{0.57} & \colhead{0.285} & \colhead{0.114} & \colhead{0.057} & \colhead{0.0114} & \colhead{0.0057} & \colhead{0.00285} & \colhead{0.00114} & \colhead{0.00057}}
\startdata
JFC & 0.0 & 0.0 & 0.0 & 0.025 & 0.75 & 1.1 & 1.6 & 1.8 & 2.6 \\
ETC-1 & 0.0 & 0.0 & 0.0 & 0.0 & 2.4 & 0.89 & 1.9 & 2.6 & 3.8 \\
ETC-2 & 0.0 & 0.0 & 0.0 & 0.0 & 0.0 & 0.0 & 0.0 & 8.7 & 57 \\
HTC-1 & 0.0 & 0.0 & 0.0 & 0.0 & 0.99 & 1.6 & 2.3 & 4.0 & 4.7 \\
HTC-2 & 0.0 & 0.0 & 0.0 & 0.0 & 0.0 & 0.0 & 0.0 & 0.0 & 0.0 \\
CTC & 0.0 & 0.0 & 0.80 & 3.3 & 8.7 & 8.2 & 9.8 & 7.6 & 10 \\
Asteroids & 0.0 & 0.0 & 0.0 & 0.10 & 0.16 & 0.0 & 0.21 & 6.8 & 84 \\
\enddata
\end{deluxetable*}

\subsection{Dust particle shape, density and SFD}   \label{subsec:Dust particle shape, density and SFD}

In previous subsection, we explained the different evolutionary tracks between the particles with different $\beta$ values. From now on, we will connect the dust SFD measured around the Earth's orbit with that measured around cometary comae.

According to measurements of the \textit{Rosetta} mission, the dust SFD at coma of 67P/Churyumov-Gerasimenko changed over time, but it appears that there are at least two bending points in SFD that can thus be approximated by a doubly broken power-law function. The bending points were determined at $\sim$10$^{-6}$g and $\sim$10$^{-4}$g for most cases except for the region around perihelion \citep{2015Sci...347a3905R, 2016ApJ...816L..32H, 2016ApJ...821...19F, 2016MNRAS.462S..78A, 2016A&A...596A..87M}. The power law exponents $\alpha$ for differential mass distributions $dn \propto m^{-\alpha} dm$ varied between $\sim$1.75$-$2.05 for the smallest masses, $\sim$0.97$-$1.67 for intermediate masses ($\sim$1.9 right after perihelion), and at approximately $\sim$2.0 for largest masses. The results from the \textit{Stardust} mission \citep{2004JGRE..10912S04G,2007ESASP.643...35G}, \textit{Giotto} mission \citep{1995A&A...304..622F}, and IR observation \citep{2010AJ....139.1491V} coincided with this point. This distribution is different from that measured around the Earth's orbit by the \textit{Pegasus} and \textit{HEOS-2} missions \citep{1985Icar...62..244G}, but the difference itself is understandable when we consider the content of the previous subsection.

As explained before, what we actually calculated were orbital evolutionary tracks as a function of initial orbits and $\beta$ values. In this subsection, at first, we converted $\beta$ to particle mass assuming particle shape and density. Next, we found the expected initial dust SFD that can explain the measured SFD around the Earth's orbit. Finally, we compared the expected initial dust SFD with the measured one and derived adequate assumptions.

Cometary dust particles have been conventionally approximated as having a compact structure and an spherical shape. For compact particles, a low mass density of 0.8 g cm$^{-3}$ was derived from modeling porous icy dust via the \textit{Rosetta} measurements \citep{2016MNRAS.462S.132F}, but a relatively wide density range of 1.9 $\pm$ 1.1 g cm$^{-3}$ was measured during the \textit{Rosetta} mission by direct comparison between cross--section and impact momenta assuming particle shape \citep{2015Sci...347a3905R}. We thus tested the three different values of particle mass densities, namely 0.8 g cm$^{-3}$, 1.9 g cm$^{-3}$, and 3.0 g cm$^{-3}$ \citep{2015Sci...347a3905R, 2016MNRAS.462S.132F}.

Despite the conventional treatment of IDPs as compact spherical particles, large fluffy aggregates (larger than hundreds of micrometers in diameter) were recently found around 67P/Churyumov-Gerasimenko via the observations of the \textit{Rosetta} spacecraft \citep{2015ApJ...802L..12F,2016Natur.537...73B,2016MNRAS.462S.304M}. Although the detailed physical properties such as the structures and/or porosity of the fluffy aggregates are not well-investigated, it is likely that these fluffy aggregates have mass densities as low as 0.001 g cm$^{-3}$ \citep{2015ApJ...802L..12F}, and fractal dimensions were estimated to be 1.87 \citep{2015ApJ...802L..12F, 2016MNRAS.462S.304M}. Under these conditions, the cross-section to mass ratio of the aggregates is expected to be constant regardless of aggregate mass, or at least exhibits only insignificant changes unlike the compact particles \citep{1992A&A...262..315M, 2016MNRAS.461.3410S}. We are not sure about the exact cause of the shallow SFD between 10$^{-6}$ $-$ 10$^{-4}$ g, but we conjecture that the fluffy aggregates that were discovered by the \textit{Rosetta} mission \citep{2015ApJ...802L..12F,2016Natur.537...73B,2016MNRAS.462S.304M} are responsible for this shallow SFD slope. The time of the change in the fluffy aggregate ratio coincides with change in SFD around perihelion \citep{2015A&A...583A..13D, 2016MNRAS.462S.210D}, and the size of large fracta from fluffy aggregates match the particle size of the smaller bending point in the SFD \citep{2016ApJ...816L..32H, 2016A&A...596A..87M}.

As the first trial, we employed the initial SFD of compact spherical dust particles with a single power law. Where $dn$ is the differential number density of dust particles of mass $m$, the initial SFD is written as $dn \propto m^{-\alpha} dm$. We tested different $\alpha$ values with intervals of 1/12. In this model, we cannot validate any initial parameters that yields less than two times the difference between observations and expectations; therefore, we tabulated the initial parameters, yielding results that are less than five times different from the observation. The derived power index, $\alpha$, is summarized in Table 4. We also show the results of the fitting in Figure 3 along with other cases. We present the ratio between our best-fit SFD and that of \citet{1985Icar...62..244G} in Figure 4.

\begin{deluxetable*}{ccccc}
\tablecaption{Best-fit parameters and expected mass supply rate for the SFD model 1\tablenotemark{a}}
\tablehead{\colhead{$\rho$ [g cm$^{-3}$]} & \colhead{$\alpha$} & \colhead{$dm_{\textrm{total}}/dt$ [t s$^{-1}$]\tablenotemark{b}} & \colhead{$dm_{\textrm{total}}/dt$ [t s$^{-1}$]\tablenotemark{c}} & \colhead{goodness of fit}\tablenotemark{d}}
\startdata
0.8 & 2.000 $-$ 2.167 & 39 $-$ 45 & 35 $-$ 43 & 0.22 $-$ 0.32\\
1.9 & 1.917 $-$ 2.000 & 39 $-$ 44 & $\sim$35 & 0.27 \\
3.0 & 1.833 $-$ 2.000 & 44 $-$ 53 & 35 $-$ 37 & 0.25 $-$ 0.27 \\
\enddata
\tablenotetext{a}{$dn \propto m^{-\alpha} dm$}
\tablenotetext{b}{Total mass supply rate into the IDP cloud complex for particles with $\beta > 0.00057$}
\tablenotetext{c}{Total mass supply rate into the IDP cloud complex for particles with $\beta > 0.00114$}
\tablenotetext{d}{Average of absolute values of log of ratio between our model and observed model \citep{1985Icar...62..244G}}
\end{deluxetable*}

\begin{figure*}
\figurenum{3}
\epsscale{1}
\plottwo{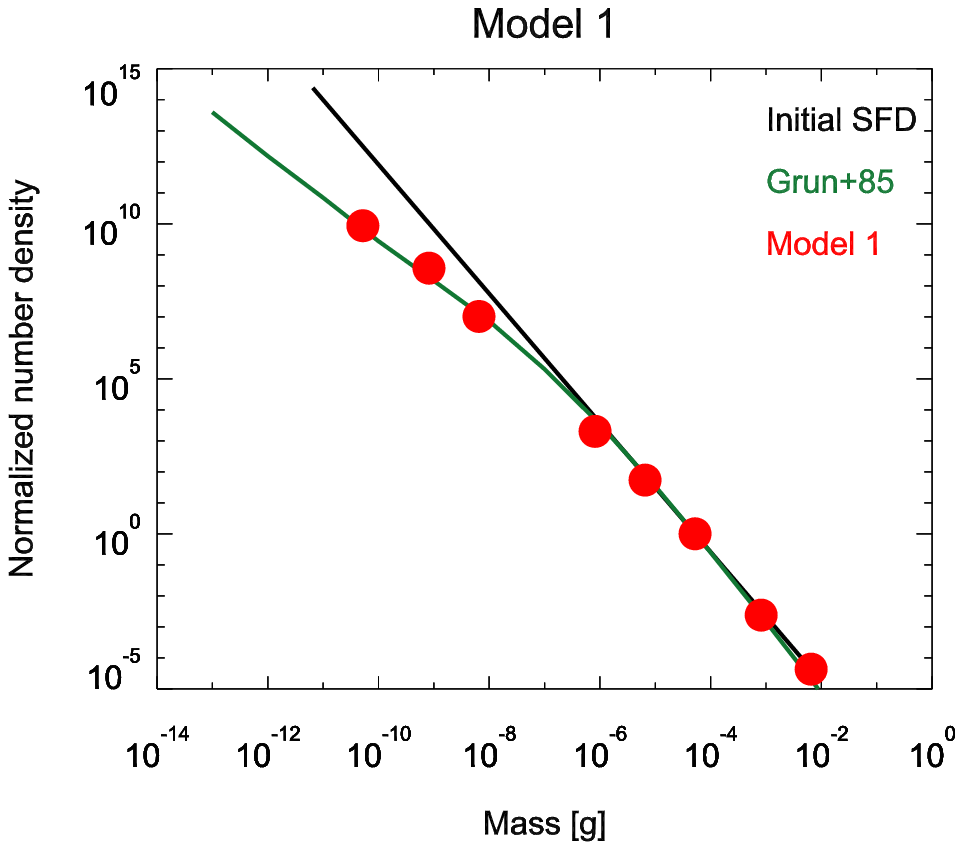}{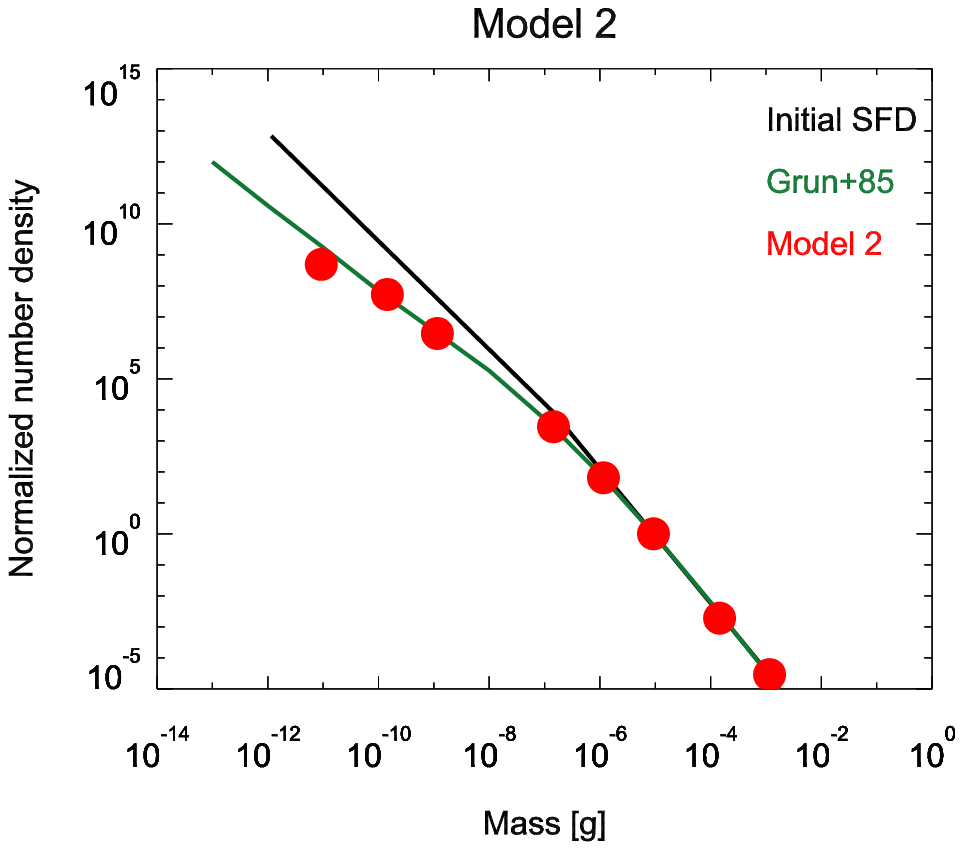}
\plottwo{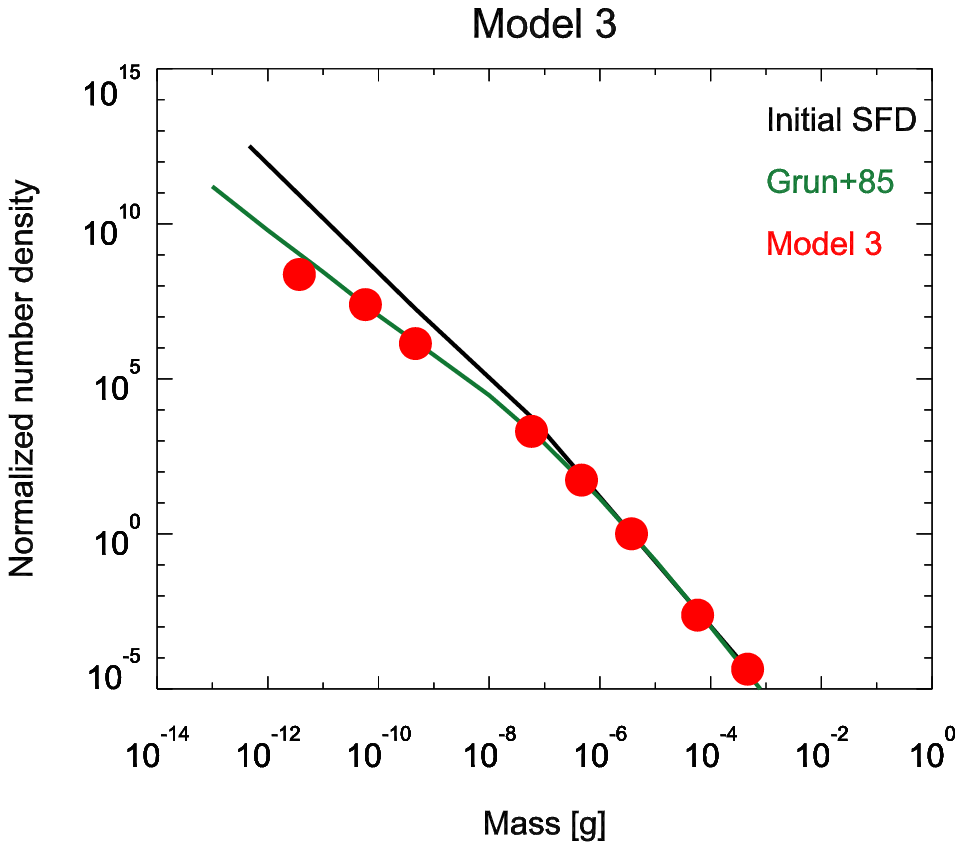}{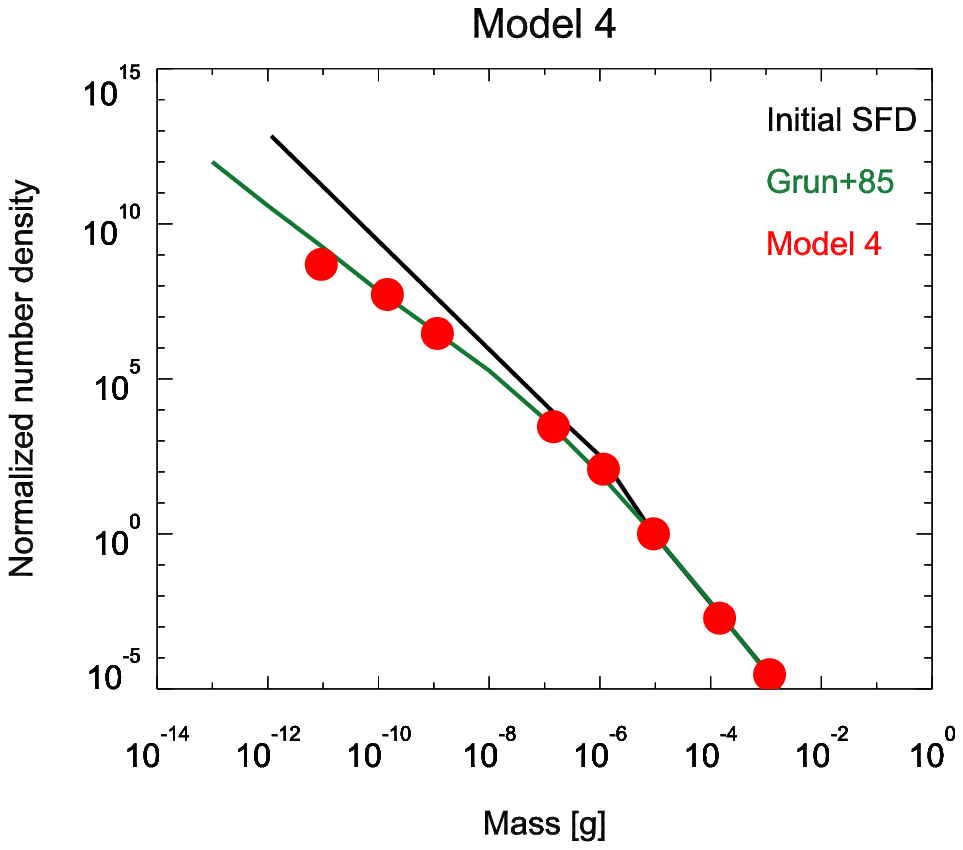}
\includegraphics[scale=0.75]{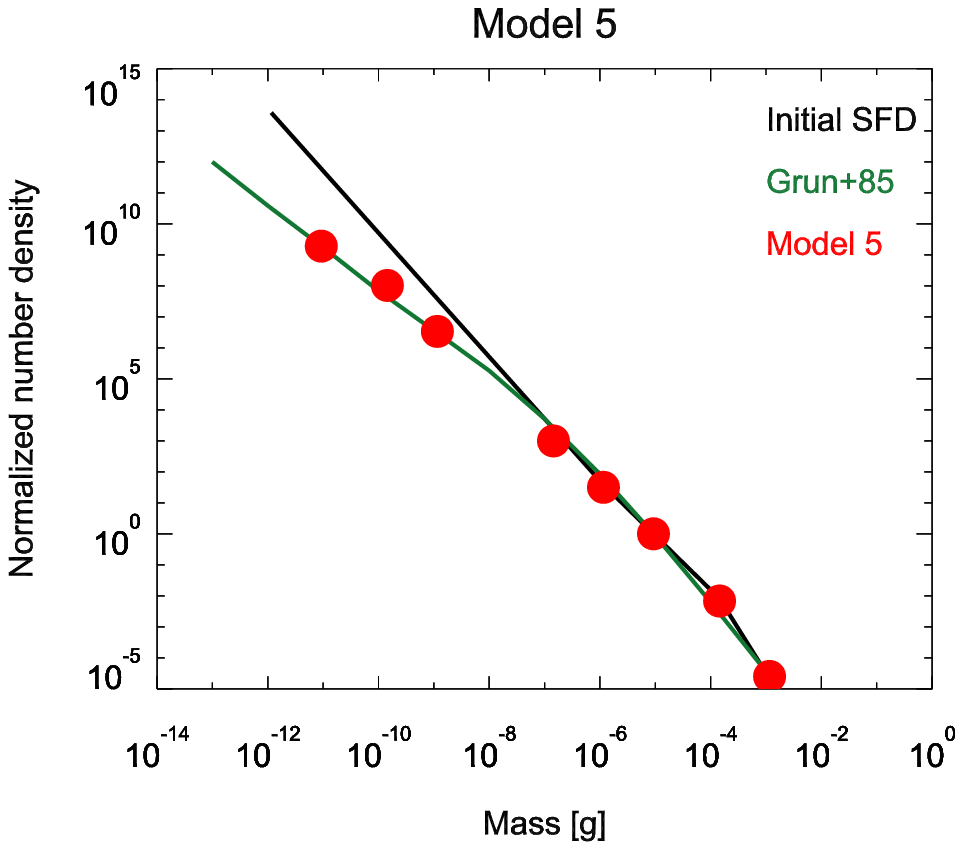}
\caption{Comparison of SFDs between the observation and our models. Black lines are initial SFD at the source regions calculated from the model parameters tabulated in Table 4$-$8. The best-fit values are used in drawing solid lines. Green lines are observed SFD around the Earth's orbit via \textit{Pegasus} and \textit{HEOS-2} spacecrafts. Red dots are SFD around the Earth's orbit derived from our models using the best fit parameters or parameters measured around cometary nuclei by \textit{Rosetta} and \textit{Stardust} missions for model 5.}
\end{figure*}

\begin{figure*}
\figurenum{4}
\epsscale{1}
\plottwo{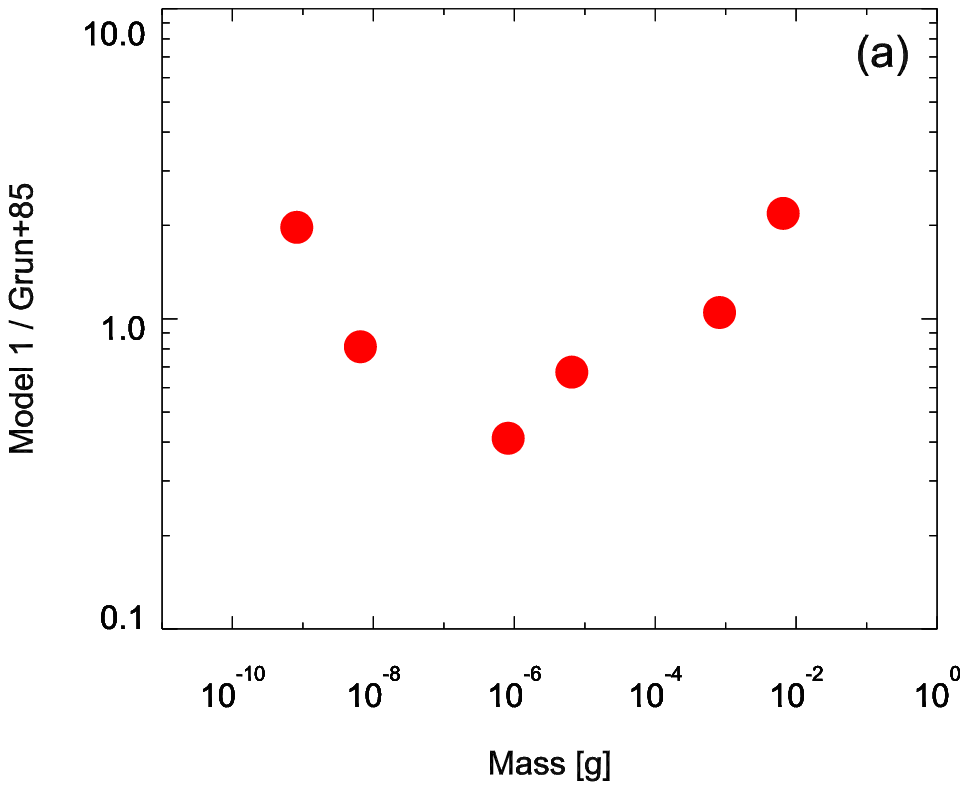}{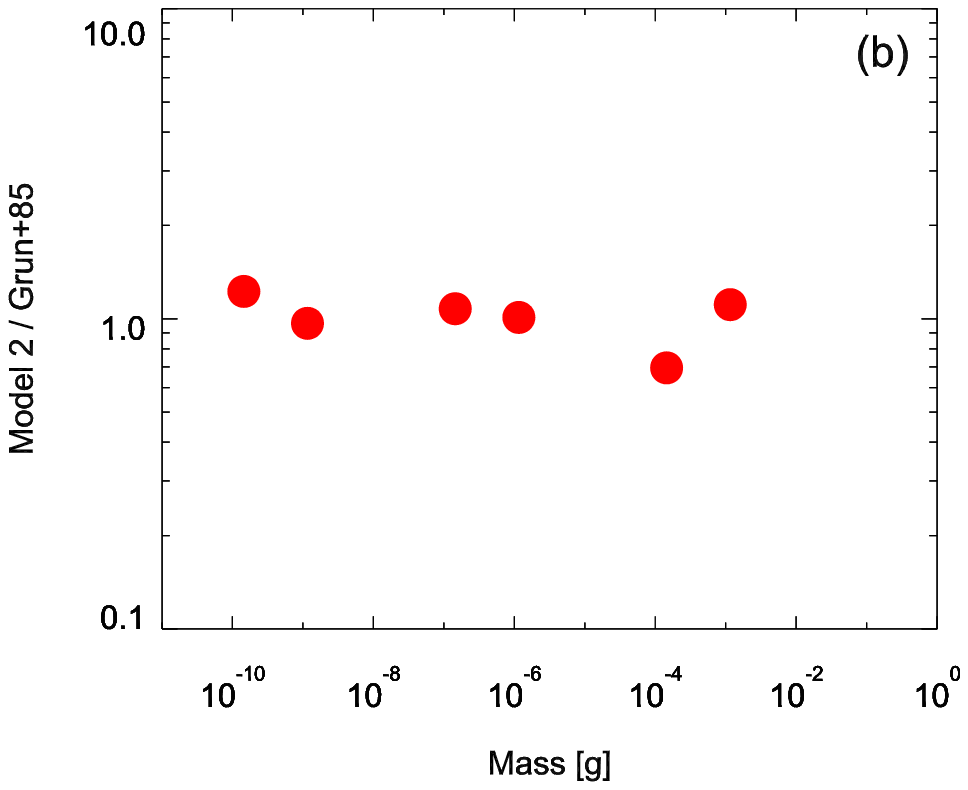}
\plottwo{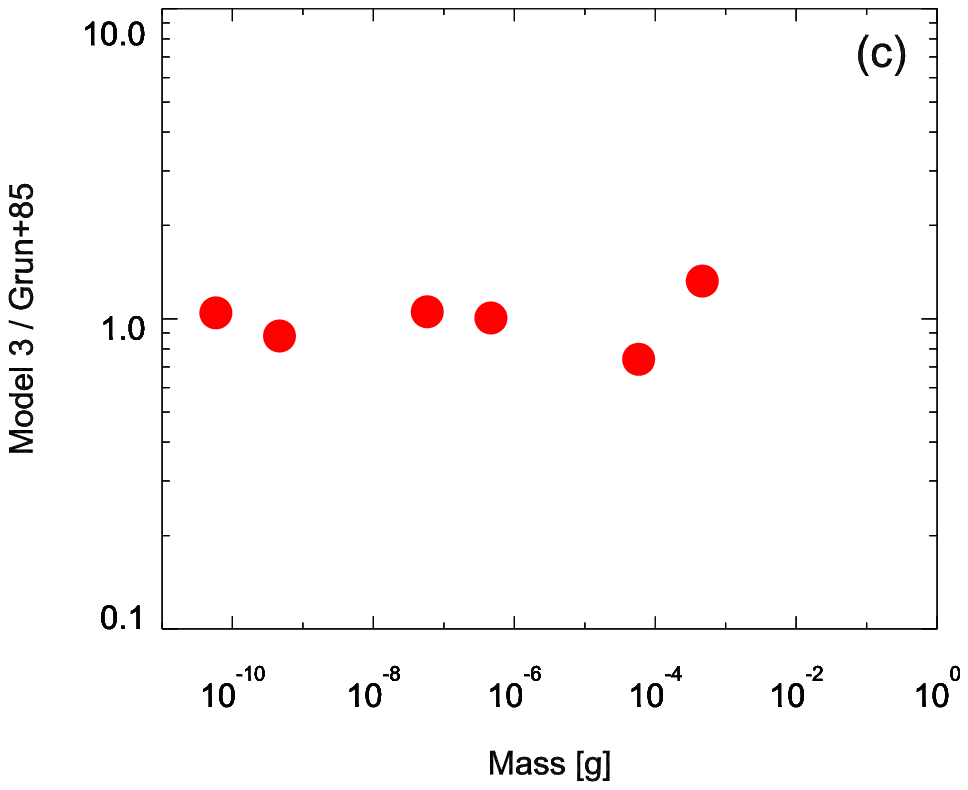}{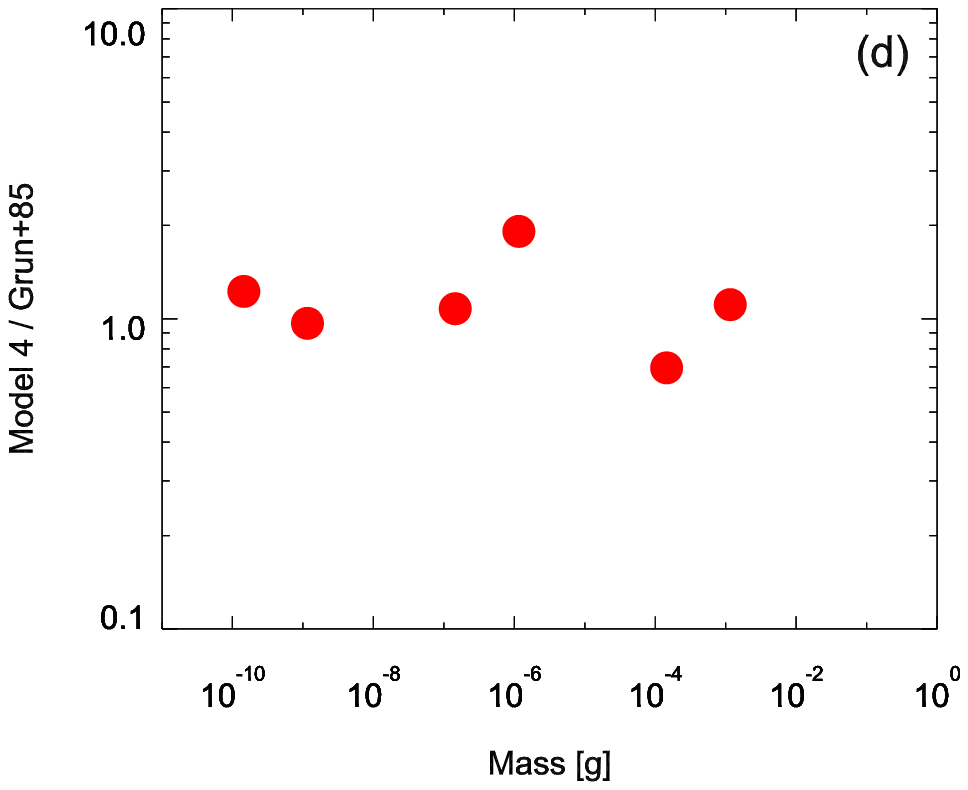}
\plottwo{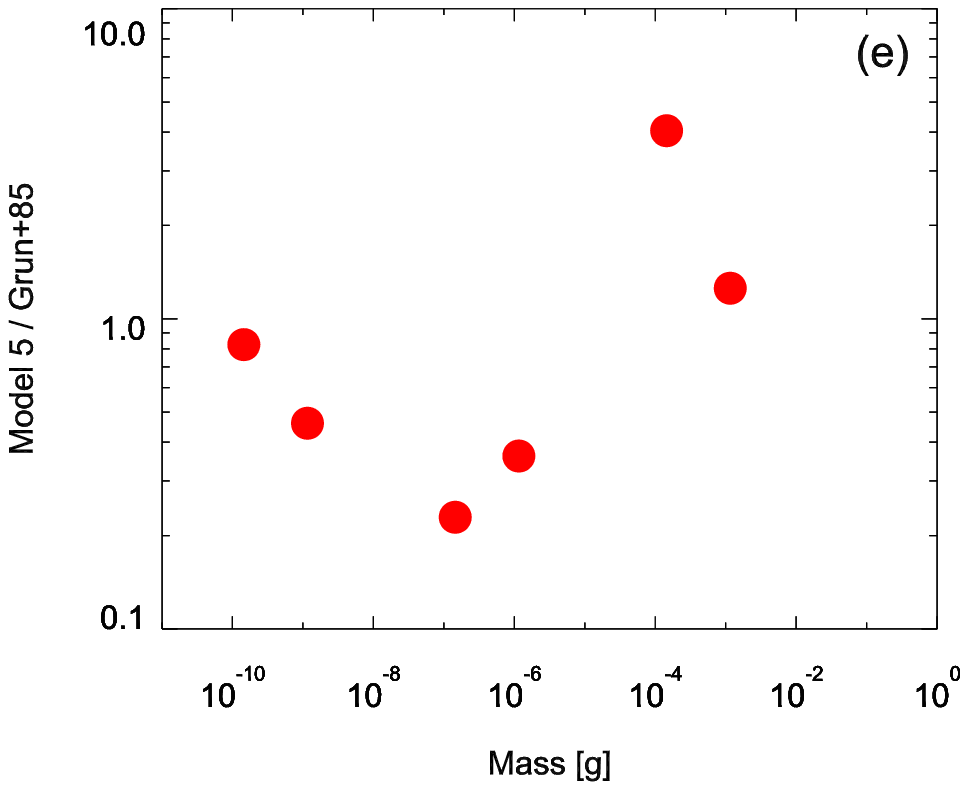}{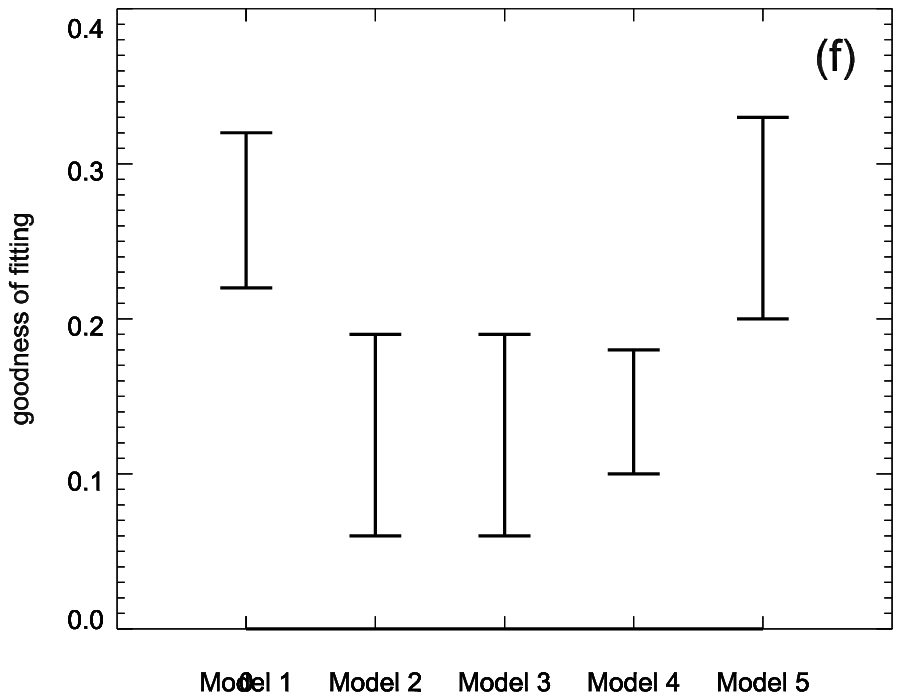}
\caption{Ratio between our best-fit SFD models and observed SFD around the Earth's orbit by \textit{Pegasus} and \textit{HEOS-2} spacecraft. Figure (e) is not from best-fit input parameters but from parameters measured around cometary nuclei by \textit{Rosetta} and \textit{Stardust} missions. (f) Goodness of fit indicates the average of absolute values of the log of the ratio between the model with the parameter and the observed model from \citep{1985Icar...62..244G}.}
\end{figure*}

Second, we employed compact spherical dust particles having a broken power-law initial SFD, namely $dn \propto m^{-\alpha_{1}} dm$ for $m \leq m_{c}$ and $dn \propto m^{-\alpha_{2}} dm$ for $m \geq m_{c}$. When fitting the data, we changed $m_{c}$ between 5 discrete $\beta$ values included in the numerical integration. The derived initial parameters are tabulated in Table 5.

\begin{deluxetable*}{ccccccc}
\tablecaption{Best-fit parameters and expected mass supply rate for the SFD model 2\tablenotemark{a}}
\tablehead{\colhead{$\rho$ [g cm$^{-3}$]} & \colhead{$m_{c}$ [g]}& \colhead{$\alpha_{1}$} & \colhead{$\alpha_{2}$} & \colhead{$dm_{\textrm{total}}/dt$ [t s$^{-1}$]\tablenotemark{b}} & \colhead{$dm_{\textrm{total}}/dt$ [t s$^{-1}$]\tablenotemark{c}} & \colhead{goodness of fit}\tablenotemark{d}}
\startdata
0.8 & 6.5 $\times$ 10$^{-9}$ & 1.583 $-$ 1.667 & 2.167 & $\sim$29 & $\sim$28 & 0.16 $-$ 0.18\\
0.8 & 8.2 $\times$ 10$^{-7}$ & 1.833 $-$ 2.000 & 2.167 $-$ 2.333 & 30 $-$ 35 & 30 $-$ 34 & 0.09 $-$ 0.15\\
0.8 & 6.5 $\times$ 10$^{-6}$ & 2.000 & 2.333 & $\sim$34 & $\sim$34 & 0.18\\
1.9 & 1.4 $\times$ 10$^{-7}$ & 1.583 $-$ 1.833 & 2.083 $-$ 2.250 & 31 $-$ 36 & 30 $-$ 32 & 0.06 $-$ 0.19\\
1.9 & 1.2 $\times$ 10$^{-6}$ & 1.833 $-$ 1.917 & 2.083 $-$ 2.250 & 34 $-$ 37 & 32 $-$ 34 & 0.11 $-$ 0.16\\
3.0 & 5.8 $\times$ 10$^{-8}$ & 1.583 $-$ 1.833 & 2.000 $-$ 2.167 & 34 $-$ 41 & 31 $-$ 34 & 0.06 $-$ 0.16\\
3.0 & 4.7 $\times$ 10$^{-7}$ & 1.750 $-$ 1.833 & 2.000 $-$ 2.167 & 36 $-$ 42 & 33 $-$ 35 & 0.11 $-$ 0.15\\
\enddata
\tablenotetext{a}{$dn \propto m^{-\alpha_{1}} dm$ for $m \leq m_{c}$, and $dn \propto m^{-\alpha_{2}} dm$ for $m \geq m_{c}$}
\tablenotetext{b}{Total mass supply rate into the IDP cloud complex for particles with $\beta > 0.00057$}
\tablenotetext{c}{Total mass supply rate into the IDP cloud complex for particles with $\beta > 0.00114$}
\tablenotetext{d}{Average of absolute values of log of ratio between our model and observed model \citep{1985Icar...62..244G}}
\end{deluxetable*}

Third, we tested the case of a doubly broken power-law SFD with compact spherical dust particles. SFD was assumed with the form of $dn \propto m^{-\alpha_{1}} dm$ for $m \leq m_{c1}$, $dn \propto m^{-\alpha_{2}} dm$ for $m_{c1} \leq m \leq m_{c2}$ and $dn \propto m^{-\alpha_{3}} dm$ for $m_{c2} \leq m$. In this case, we assumed $\alpha_{1} > \alpha_{2}$, $\alpha_{3} > \alpha_{2}$ and limited parameters as in the previous case. The results are tabulated in Table 6.

\begin{deluxetable*}{ccccccccc}
\tablecaption{Best-fit parameters and expected mass supply rate for the SFD model 3\tablenotemark{a}}
\tablehead{\colhead{$\rho$} & \colhead{$m_{c1}$} & \colhead{$m_{c2}$} & \colhead{$\alpha_{1}$} & \colhead{$\alpha_{2}$} & \colhead{$\alpha_{3}$} & \colhead{$dm_{\textrm{total}}/dt$\tablenotemark{b}} & \colhead{$dm_{\textrm{total}}/dt$\tablenotemark{c}} & \colhead{goodness of}\\
\colhead{[g cm$^{-3}$]} & \colhead{[g]} & \colhead{[g]} & \colhead{} & \colhead{} & \colhead{} & \colhead{[t s$^{-1}$]} & \colhead{[t s$^{-1}$]} & \colhead{fit}\tablenotemark{d}}
\startdata
0.8 & 6.5$\times$10$^{-9}$ & 8.2$\times$10$^{-7}$ $-$ 6.5$\times$10$^{-6}$ & 1.917 $-$ 2.250 & 1.833 $-$ 2.000 & 2.167 $-$ 2.333 & 31 $-$ 46 & 31 $-$ 45 & 0.10 $-$ 0.19\\
1.9 & 6.5$\times$10$^{-9}$ $-$ 8.2$\times$10$^{-7}$ & 8.2$\times$10$^{-7}$ $-$ 6.5$\times$10$^{-6}$ & 1.667 $-$ 2.167 & 1.583 $-$ 1.833 & 2.083 $-$ 2.250 & 32 $-$ 39 & 30 $-$ 36 & 0.07 $-$ 0.19\\
3.0 & 6.5$\times$10$^{-9}$ $-$ 8.2$\times$10$^{-7}$ & 8.2$\times$10$^{-7}$ $-$ 6.5$\times$10$^{-6}$ & 1.667 $-$ 2.250 & 1.583 $-$ 1.833 & 2.000 $-$ 2.167 & 34 $-$ 44 & 31 $-$ 38 & 0.06 $-$ 0.16\\
\enddata
\tablenotetext{a}{$dn \propto m^{-\alpha_{1}} dm$ for $m \leq m_{c1}$, $dn \propto m^{-\alpha_{2}} dm$ for $m_{c1} \leq m \leq m_{c2}$ and $dn \propto m^{-\alpha_{3}} dm$ for $m_{c2} \leq m$}
\tablenotetext{b}{total mass supply rate into the IDP cloud complex by particles with $\beta > 0.00057$}
\tablenotetext{c}{total mass supply rate into the IDP cloud complex by particles with $\beta > 0.00114$}
\tablenotetext{d}{average of absolute values of log of ratio between our model and observed model \citep{1985Icar...62..244G}}
\end{deluxetable*}

The next model is composed of both compact spherical particles and fluffy aggregates. We assumed that the fluffy aggregates have constant $\beta$ values regardless of mass \citep{1992A&A...262..315M, 2016MNRAS.461.3410S}. The SFD of compact spherical particles was assumed to be $dn \propto m^{-\alpha_{1}} dm$ for $m \leq m_{c_{1}}$ and $dn \propto m^{-\alpha_{3}} dm$ for $m \geq m_{c_{1}}$. The SFD of fluffy aggregates was assumed to be the sum of fluffy aggregates, and compact spherical particles become $dn \propto m^{-\alpha_{2}} dm$ between masses of $m_{c_{1}}$ and $m_{c_{2}}$. The results are presented in Table 7.

\begin{deluxetable*}{ccccccccccc}
\tablecaption{Best-fit parameters, expected mass supply rate and contribution of fluffy particles in zodiacal light for the SFD model 4\tablenotemark{a}}
\tablehead{\colhead{$\rho$} & \colhead{$m_{c1}$} & \colhead{$m_{c2}$} & \colhead{$\alpha_{1}$} & \colhead{$\alpha_{2}$} & \colhead{$\alpha_{3}$} & \colhead{$dm_{\textrm{comp}}/dt$\tablenotemark{b}} & \colhead{$dm_{\textrm{fluf}}/dt$\tablenotemark{c}} & \colhead{$ZL_{\textrm{fluf}}/ZL$\tablenotemark{d}} & \colhead{$m_{\textrm{fluf}}/m_{\textrm{total}}$\tablenotemark{e}} & \colhead{goodness of}\\
\colhead{[g cm$^{-3}$]} & \colhead{[g]} & \colhead{[g]} & \colhead{} & \colhead{} & \colhead{} & \colhead{[t s$^{-1}$]} & \colhead{[t s$^{-1}$]} & \colhead{} & \colhead{} & \colhead{fit\tablenotemark{f}}}
\startdata
0.8 & 8.2$\times$10$^{-7}$ & 6.5$\times$10$^{-6}$ & 1.917$-$2.000 & 1.583$-$1.917 & 2.167$-$2.250 & 32$-$35 & 0.05$-$0.2 & 0.005 & 0.003$-$0.007 & 0.12$-$0.18\\
1.9 & 1.5$\times$10$^{-7}$ & 1.2$\times$10$^{-6}$ & 1.750$-$1.833 & 1.583$-$1.750 & 2.083$-$2.167 & 33$-$36 & 0.1$-$0.3 & 0.008 & 0.004$-$0.006 & 0.10$-$0.14\\
3.0 & 5.8$\times$10$^{-8}$ & 4.7$\times$10$^{-7}$ & 1.667$-$1.833 & 1.583$-$1.750 & 2.000$-$2.083 & 36$-$41 & 0.1$-$0.3 & 0.007 & 0.002$-$0.006 & 0.10$-$0.17\\
\enddata
\tablenotetext{a}{Explanations of functional form in text}
\tablenotetext{b}{Total mass supply rate into the IDP cloud complex for compact particles}
\tablenotetext{c}{Total mass supply rate into the IDP cloud complex for fluffy aggregates}
\tablenotetext{d}{Fraction of gegenschein brightness contributed by fluffy aggregates}
\tablenotetext{e}{Fraction of fluffy aggregates mass around the Earth's orbit}
\tablenotetext{f}{Average of absolute values of log of ratio between our model and observed model \citep{1985Icar...62..244G}}
\end{deluxetable*}

The best-fit initial SFD of previous paragraph has bending points at smaller mass than SFD measured at cometary comae, and the slopes steeper than measurements in $0.0057 < \beta < 0.000114$ (Table 7). As a final model, we applied SFD with the same functional form as that in the above model but fixed $m_{c1}$ and $m_{c2}$ around the observed values. Then, the $\alpha_{1}$, $\alpha_{2}$, and $\alpha_{3}$ values will be near the observed values, especially when compact particle density is higher than 1.9 g cm$^{-3}$. In those cases, the ratio between our estimation and \citet{1985Icar...62..244G}'s model worsened by a factor of three for some sizes. However, our model using this observed input parameter value is still not far from the observed SFD in the Earth's orbit as presented in Table 8 and Figures 3 and 4.

\begin{deluxetable*}{ccccccccccc}
\tablecaption{Best-fit parameters, expected mass supply rate and contribution of fluffy particles in zodiacal light for model 5\tablenotemark{a}}
\tablehead{\colhead{$\rho$} & \colhead{$m_{c1}$} & \colhead{$m_{c2}$} & \colhead{$\alpha_{1}$} & \colhead{$\alpha_{2}$} & \colhead{$\alpha_{3}$} & \colhead{$dm_{\textrm{comp}}/dt$\tablenotemark{b}} & \colhead{$dm_{\textrm{fluf}}/dt$\tablenotemark{c}} & \colhead{$ZL_{\textrm{fluf}}/ZL$\tablenotemark{d}} & \colhead{$m_{\textrm{fluf}}/m_{\textrm{total}}$\tablenotemark{e}} & \colhead{goodness of}\\
\colhead{[g cm$^{-3}$]} & \colhead{[g]} & \colhead{[g]} & \colhead{} & \colhead{} & \colhead{} & \colhead{[t s$^{-1}$]} & \colhead{[t s$^{-1}$]} & \colhead{} & \colhead{} & \colhead{fit\tablenotemark{f}}}
\startdata
0.8 & 6.5$\times$10$^{-6}$ & 8.2$\times$10$^{-4}$ & 2.083$-$2.167 & 1.833$-$2.000 & 2.083$-$2.250 & 34$-$41 & 0.8$-$3 & 0.07 & 0.12$-$0.25 & 0.20$-$0.33\\
1.9 & 1.2$\times$10$^{-6}$ & 1.5$\times$10$^{-4}$ & 1.917$-$2.000 & 1.583$-$1.750 & 1.917$-$2.250 & 25$-$36 & 4$-$8 & 0.20 & 0.17$-$0.37 & 0.21$-$0.33\\
3.0 & 3.7$\times$10$^{-6}$ & 5.8$\times$10$^{-5}$ & 1.833$-$1.917 & 1.667$-$1.833 & 1.917$-$2.333 & 35$-$48 & 0.06$-$0.5 & 0.007 & 0.17$-$0.44 & 0.20$-$0.31\\
\enddata
\tablenotetext{a}{Explanations of functional form in text}
\tablenotetext{b}{Total mass supply rate into the IDP cloud complex for compact particles}
\tablenotetext{c}{Total mass supply rate into the IDP cloud complex for fluffy aggregates}
\tablenotetext{d}{Fraction of gegenschein brightness caused by fluffy aggregates}
\tablenotetext{e}{Fraction of fluffy aggregates mass around the Earth's orbit}
\tablenotetext{f}{Average of absolute values of log of ratio between our model and observed model \citep{1985Icar...62..244G}}
\end{deluxetable*}

As presented in Figure 4 and Tables 4--8, different dust ejection conditions for dust particle shape and density can be validated with the appropriate initial SFD. It is noteworthy that the dust ejection conditions measured around 67P/Churyumov-Gerasimenko and 81P/Wild 2 concur with dust particle measurements around the Earth's orbit when we assume the existence of fluffy aggregates of discovered size range. We consider that the various dust environments measured around cometary nuclei are in accordance with our knowledge about the IDP cloud complex and its cometary origin and that the target comets of space missions (i.e., 1P, 67P, and 81P) are not extraordinary but may represent ordinary dust ejections from comets.

\subsection{Mass supply rate to the IDP cloud complex}   \label{subsec:Mass supply rate to the IDP cloud complex}

By scaling our model SFD by the observed gegenschein brightness, we calculated the expected mass supply rate to the IDP cloud complex. As presented in Tables 4$-$8 and Figure 5, the total dust supply rate does not vary greatly according to dust particle shape, density and the initial SFD when the latter is confined by the observed SFD around the Earth's orbit and gegenschein brightness. The required dust supply rate for particles with $\beta >$ 0.00057 was 39$-$45 tons s$^{-1}$ for model 1, 29$-$41 tons s$^{-1}$ for model 2, 31$-$49 tons s$^{-1}$ for model 3, 32$-$ 41 tons s$^{-1}$ for model 4, and 32$-$43 tons s$^{-1}$ for model 5. We thus find that the different SFD models in this research do not significantly change the mass supply rate. This value is roughly consistent with but slightly larger than \citet{2011ApJ...743..129N}'s estimation of 1.6$-$25 t s$^{-1}$. When we consider that our derived SFD is shallower than \citet{2011ApJ...743..129N}'s, we believe that this difference is understandable.

Even though we did not include dust particles larger than $\beta = 0.00057$, we expect that this omission will not critically change our conclusion because the mass supply rates for $\beta \geq 0.00114$ and $\beta \geq 0.00057$ are not significantly different from each other, as shown in Tables 4$-$8. 
\begin{figure}
\figurenum{5}
\epsscale{1}
\plotone{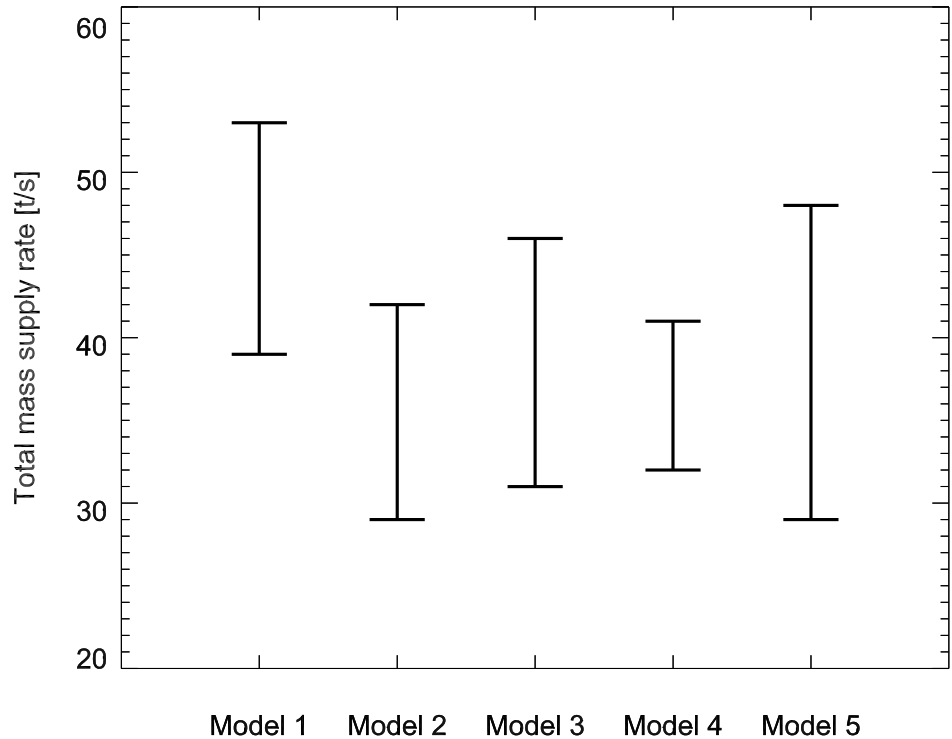}
\caption{Total mass supply rate for models with validated initial parameters. The resulting values from all different densities are included.}
\end{figure}

\subsection{Mutual collisions}   \label{subsec:Mutual collisions}

Collisional probabilities of the solar system objects have been investigated since the pioneering works by \citet{1951PRIA...54..165O} and \citet{1969JGR....74.2531D}. Collisional lifetime of IDPs was estimated under the assumption of the fixed circular orbits \citep{1978codu.book..527D, 1985Icar...62..244G, 1986MNRAS.218..185S}. It is noticed that catastrophic collisions among IDPs could be a dominant mechanism for loosing IDPs if their particle sizes are larger than $\sim$ 200 $\mu$m \citep{1978codu.book..527D, 1985Icar...62..244G, 1986MNRAS.218..185S}. Despite the simplified models in these references where IDPs revolve in the fixed circular orbits, their estimated values have been considered in recent dynamical studies for cometary dust particles with eccentric orbits \citep{2009Icar..201..295W, 2011ApJ...743..129N, 2014ApJ...789...25P}. More recently, \citet{2016pimo.conf..284S} derived the collisional lifetime for particles on various fixed eccentric orbits, and the lifetime was comparable with P-R lifetime for 100 $\mu$m particles. On another front, the orbital distribution of helion meteors favors a collisional lifetime longer than \citet{1985Icar...62..244G}'s estimate \citep{2011ApJ...743..129N}. From the observational aspect, \citet{2016Icar..266..384J} discovered 7-mm meteorites delivered to the Earth which have low eccentric orbits, suggesting that such dust particles were migrated via P--R drag before collisional breakup. To sum up these bibliographic background, it is required to calculate the collisional lifetime in a more realistic manner, where the orbits of IDPs evolve under planetary perturbations and P--R drag.

As mentioned in section 3.1, most dust particles from JFCs are kicked out to orbits with large $a$ values and stay for a long time in the outer solar system. Frequently, even the particles eventually fall to the Sun, are kicked out to orbits with large $a$ values and spent most of their lifetimes in such orbits. Furthermore, the eccentricity of their orbits does not decrease as in the situation with no planets. When we consider these factors along with the dynamical lifetimes of dust particles that are shorter than the P--R timescale, we can expect a lower probability of mutual collision in the IDP cloud complex. In this paper, we quantitatively treated the mutual collisions and discussed whether the impact effect is significant. We did not consider the impact velocity and consequence of collision, but calculated the probability of collision regardless of impact velocity. We think this choice can simplify required treatments during calculation.

We calculated the volume swept by a dust particle during orbital evolution. At each epoch, we also calculated the cross-section density on the positions of the dust particle orbit as a function of particle size. By integrating the multiplication of these two values (i.e., the swept volume and the cross-section density), we derived the cross-section swept by the dust particle. By dividing the swept cross-section by the cross-section of responsible single particle (i.e., the projectile), we derived the contact probability for given epochs. We defined the contact timescale by dividing the dynamical lifetime by total contact probability, which is the integration of the contact probability over the entire dynamical lifetime. The contact timescale was calculated as a function of the smallest projectile size considered. The timescales are shown in Figure 2.

As we can see in Figure 2, mutual contact between particles with a size ratio that is larger than 1:100 (a mass ratio of one to million) will not happen in the IDP cloud complex. Because we did not concern the impact speed of these contacts, we are not sure how large a projectile is required to break the target dust particle, but it is expected that a projectile 100 times smaller than the target will not destroy it. In addition, we do not know the projectile size that is required to break the fluffy aggregates. The values in Figure 2 are from the case in which fluffy aggregates have the longest dynamical lifetime and the largest size (a 3-mm diameter), the case in which particles most easily experience collision. When we consider that fluffy aggregates were fragmented by collisions with impact speeds of $\sim$ cm s$^{-1}$ in the COSIMA detector on the \textit{Rosetta} spacecraft \citep{2016ApJ...816L..32H}, we think contact with a 25-$\mu$m projectile would break the 3-mm fluffy aggregates at least partially, but we are not sure about exact consequences of contact. However, in overall, we can see that fluffy aggregates will not experience mutual collisions, especially at sizes smaller than millimeters in diameter.

\section{The relative importance of particles: source, size, and shape}   \label{sec:The relative importance of particles: source, size, and shape}

We list the relative contribution to the gegenschein brightness as a function of source and size in Table 9, adopting the case of a compact particle model with a 1.9-g cm$^{-3}$ density. The difference is not large for other cases. A dominant fraction ($>$ 90\%) of gegenschein brightness is caused by dust particle from JFCs, the $\beta$ of which is in the range of 0.057 $-$ 0.00285 (10 $-$ 200 $\mu$m in diameter for $\rho = 1.9$ g cm$^{-3}$). However, when we derived the relative contributions of each source to the dust particles around the orbit of the Earth as a function of size, for the particles $\gtrsim$ 1 mm, the contribution of ETC-2 dust particles is even larger than that of JFC particles, as shown in Table 10. Typically, dust particles ejected from 2P/Encke have lifetimes of $\sim$100 thousands years even for the largest particles, requiring $\sim$half of the lifetime to reach $a < 1$ au. Although the past activities of 2P/Encke and similar objects (i.e., ETC--2) are not well known, we found that dust ejected from 2P/Encke will be more important than dust ejected from any other comet at least for large dust particles in the inner solar system. Additionally, large dust particles ejected from AAs would have larger contribution to the IDP cloud complex at small heliocentric distances than to the dust supply rate, though further research is needed to survey the dust production rates and the occurrence of AAs. Therefore, when we observe zodiacal light, meteors in the sky, and meteorites on the surface of the Earth, there is the possibility that we are observing different types of particles for each observation.

\begin{deluxetable*}{ccccc}
\tablecaption{Relative contribution to the total gegenschein brightness}
\tablehead{\colhead{$\beta$} & \colhead{JFC} & \colhead{HTC} & \colhead{ETC} & \colhead{AA}}
\startdata
0.057 $-$ & 12.2 & 0.0 & 0.8 & 0.7 \\
0.0114 $-$ 0.057 & 36.9 & 0.0 & 1.6 & 3.3 \\
0.00285 $-$ 0.011 & 32.6 & 0.0 & 1.0 & 4.0 \\
$-$ 0.00285 & 4.9 & 0.0 & 0.1 & 1.9 \\
\enddata
\end{deluxetable*}

\begin{deluxetable*}{cccccccc}
\tablecaption{Relative contribution to the number density around the Earth's orbit. Percentage calculated size by size}
\tablehead{\colhead{$\beta = $} & \colhead{0.114} & \colhead{0.057} & \colhead{0.0114} & \colhead{0.0057} & \colhead{0.00285} & \colhead{0.00114} & \colhead{0.00057}}
\startdata
JFC & 87.6 & 89.1 & 81.1 & 85.5 & 81.0 & 52.9 & 37.5\\
ETC-1 & 6.0 & 4.2 & 3.6 & 2.8 & 2.2 & 0.4 & 0.9\\
ETC-2 & 6.4 & 6.7 & 15.2 & 11.7 & 16.8 & 46.7 & 61.6\\
\enddata
\end{deluxetable*}

We tabulated following values for compact particles and fluffy aggregates in Tables 7 and 8: mass supply rate, contribution to the gegenschein brightness, and fractional mass around the Earth's orbit. As measured by \citet{2015ApJ...802L..12F} in the coma of 67P/Churyumov-Gerasimenko, the fluffy aggregates have small relative contributions at the moment of ejection. The contribution of these fluffy aggregates to the zodiacal light brightness is also small, that is, as much as $\sim$1$-$20 \%. Therefore, in our methodology, the total mass supply rate to the IDP cloud complex does not largely change with or without fluffy aggregates. However, as shown in Table 8, their contribution may be significantly large in the sense of number density (or mass) near the Earth's orbit. Therefore, our results suggest that fluffy IDPs discovered in the Earth's stratosphere, which have similar fractal dimensions as the fluffy aggregates from 67P/Churyumov-Gerasimenko \citep{KATYAL2014290}, were possibly ejected from comets such as 67P. 

\section{Summary}   \label{sec:Summary}

In this study, we followed the evolution of dust particles ejected from cometary nuclei, considering different initial orbits, shapes and SFDs. According to our model, the total mass supply rate from comets to the IDP cloud complex is 29$-$53 t s$^{-1}$. We discovered that if we introduce fluffy aggregates, the dust SFD measured in cometary comae can be evolved into the dust SFD measured around the Earth's orbit. Even though initial fractional mass of fluffy aggregates is small at the source regions, cometary comae, these aggregates may represent a relatively large fraction of the mass around the Earth's orbit. Based on our findings above, we conjecture an initial dust SFD and mass supply rate that are different from those in \citet{2009Icar..201..295W,2011ApJ...743..129N} without altering their conclusions of the dominant cometary contribution to IDPs, zodiacal light brightness model and helion meteor orbital distribution.

Furthermore, we identified that large dust particles ejected from JFCs cannot be easily transferred to a small $a$ orbit because of close encounters with Jupiter. Therefore, the contribution from ETC-2s may be high for large dust particles in small $a$ orbits.

Finally, we directly calculated the probability of mutual collisions between dust particles in the IDP cloud complex and concluded that mutual collisions are ignorable. This collisional lifetime is figurally (a factor of 10 or less) longer than the simplified estimate by \citet{2016pimo.conf..284S}, while the dynamical lifetime is more than an order of magnitude shorter than the collisional lifetime in \citet{2016pimo.conf..284S}. We conjecture that \citet{2016Icar..266..384J}'s discovery, large meteorites on low eccentricity orbits, were possibly caused by fluffy aggregates, which have small eccentricities around the Earth's orbit, unlike the compact particles from 2P/Encke or JFCs.

\ \\

\acknowledgments We thank the reviewer, Jeremie Vaubaillon, for his constructive comments which help us to improve this manuscript. This work was supported by the National Research Foundation of Korea (NRF) funded by the South Korean government (MEST) (Grant No. 2015R1D1A1A01060025).

\bibliographystyle{aasjournal}
\bibliography{Draft_Com_Sim}

\begin{thebibliography}{}
\expandafter\ifx\csname natexlab\endcsname\relax\def\natexlab#1{#1}\fi

\bibitem[{{Agarwal} {et~al.}(2016){Agarwal}, {A'Hearn}, {Vincent},
  {G{\"u}ttler}, {H{\"o}fner}, {Sierks}, {Tubiana}, {Barbieri}, {Lamy},
  {Rodrigo}, {Koschny}, {Rickman}, {Barucci}, {Bertaux}, {Bertini},
  {Boudreault}, {Cremonese}, {Da Deppo}, {Davidsson}, {Debei}, {De Cecco},
  {Deller}, {Fornasier}, {Fulle}, {Gicquel}, {Groussin}, {Guti{\'e}rrez},
  {Hofmann}, {Hviid}, {Ip}, {Jorda}, {Keller}, {Knollenberg}, {Kramm},
  {K{\"u}hrt}, {K{\"u}ppers}, {Lara}, {Lazzarin}, {Lopez Moreno}, {Marzari},
  {Naletto}, {Oklay}, {Shi}, \& {Thomas}}]{2016MNRAS.462S..78A}
{Agarwal}, J., {A'Hearn}, M.~F., {Vincent}, J.-B., {et~al.} 2016, \mnras, 462,
  S78

\bibitem[{{Bentley} {et~al.}(2016){Bentley}, {Schmied}, {Mannel}, {Torkar},
  {Jeszenszky}, {Romstedt}, {Levasseur-Regourd}, {Weber}, {Jessberger},
  {Ehrenfreund}, {Koeberl}, \& {Havnes}}]{2016Natur.537...73B}
{Bentley}, M.~S., {Schmied}, R., {Mannel}, T., {et~al.} 2016, \nat, 537, 73

\bibitem[{{Bradley}(2003)}]{2003TrGeo...1..689B}
{Bradley}, J.~P. 2003, Treatise on Geochemistry, 1, 689

\bibitem[{{Burns} {et~al.}(1979){Burns}, {Lamy}, \&
  {Soter}}]{1979Icar...40....1B}
{Burns}, J.~A., {Lamy}, P.~L., \& {Soter}, S. 1979, \icarus, 40, 1

\bibitem[{{Chambers}(1999)}]{1999MNRAS.304..793C}
{Chambers}, J.~E. 1999, \mnras, 304, 793

\bibitem[{{Della Corte} {et~al.}(2015){Della Corte}, {Rotundi}, {Fulle},
  {Gruen}, {Weissman}, {Sordini}, {Ferrari}, {Ivanovski}, {Lucarelli},
  {Accolla}, {Zakharov}, {Mazzotta Epifani}, {Lopez-Moreno}, {Rodriguez},
  {Colangeli}, {Palumbo}, {Bussoletti}, {Crifo}, {Esposito}, {Green}, {Lamy},
  {McDonnell}, {Mennella}, {Molina}, {Morales}, {Moreno}, {Ortiz}, {Palomba},
  {Perrin}, {Rietmeijer}, {Rodrigo}, {Zarnecki}, {Cosi}, {Giovane},
  {Gustafson}, {Herranz}, {Jeronimo}, {Leese}, {Lopez-Jimenez}, \&
  {Altobelli}}]{2015A&A...583A..13D}
{Della Corte}, V., {Rotundi}, A., {Fulle}, M., {et~al.} 2015, \aap, 583, A13

\bibitem[{{Della Corte} {et~al.}(2016){Della Corte}, {Rotundi}, {Fulle},
  {Ivanovski}, {Green}, {Rietmeijer}, {Colangeli}, {Palumbo}, {Sordini},
  {Ferrari}, {Accolla}, {Zakharov}, {Mazzotta Epifani}, {Weissman}, {Gruen},
  {Lopez-Moreno}, {Rodriguez}, {Bussoletti}, {Crifo}, {Esposito}, {Lamy},
  {McDonnell}, {Mennella}, {Molina}, {Morales}, {Moreno}, {Palomba}, {Perrin},
  {Rodrigo}, {Zarnecki}, {Cosi}, {Giovane}, {Gustafson}, {Ortiz}, {Jeronimo},
  {Leese}, {Herranz}, {Liuzzi}, \& {Lopez-Jimenez}}]{2016MNRAS.462S.210D}
---. 2016, \mnras, 462, S210

\bibitem[{{Dohnanyi}(1969)}]{1969JGR....74.2531D}
{Dohnanyi}, J.~S. 1969, \jgr, 74, 2531

\bibitem[{{Dohnanyi}(1978)}]{1978codu.book..527D}
---. 1978, {Particle dynamics}, ed. J.~A.~M. {McDonnell}, 527--605

\bibitem[{{Everhart}(1985)}]{1985dcto.proc..185E}
{Everhart}, E. 1985, in Dynamics of Comets: Their Origin and Evolution,
  Proceedings of IAU Colloq. 83, held in Rome, Italy, June 11-15, 1984. Edited
  by Andrea Carusi and Giovanni B. Valsecchi. Dordrecht: Reidel, Astrophysics
  and Space Science Library. Volume 115, 1985, p.185, ed. A.~{Carusi} \& G.~B.
  {Valsecchi}, 185

\bibitem[{{Fechtig} {et~al.}(2001){Fechtig}, {Leinert}, \&
  {Berg}}]{2001InterpD..Fech}
{Fechtig}, H., {Leinert}, C., \& {Berg}, O.~E. 2001, Interplanetary Dust, ed.
  E.~{Grun}, B.~A.~S. {Gustafson}, S.~{Dermott}, \& H.~{Fechtig} (Springer
  Berlin Heidelberg), 1--56

\bibitem[{{Fern{\'a}ndez} {et~al.}(2013){Fern{\'a}ndez}, {Kelley}, {Lamy},
  {Toth}, {Groussin}, {Lisse}, {A'Hearn}, {Bauer}, {Campins}, {Fitzsimmons},
  {Licandro}, {Lowry}, {Meech}, {Pittichov{\'a}}, {Reach}, {Snodgrass}, \&
  {Weaver}}]{2013Icar..226.1138F}
{Fern{\'a}ndez}, Y.~R., {Kelley}, M.~S., {Lamy}, P.~L., {et~al.} 2013, \icarus,
  226, 1138

\bibitem[{{Fulle} {et~al.}(1995){Fulle}, {Colangeli}, {Mennella}, {Rotundi}, \&
  {Bussoletti}}]{1995A&A...304..622F}
{Fulle}, M., {Colangeli}, L., {Mennella}, V., {Rotundi}, A., \& {Bussoletti},
  E. 1995, \aap, 304, 622

\bibitem[{{Fulle} {et~al.}(2015){Fulle}, {Della Corte}, {Rotundi}, {Weissman},
  {Juhasz}, {Szego}, {Sordini}, {Ferrari}, {Ivanovski}, {Lucarelli}, {Accolla},
  {Merouane}, {Zakharov}, {Mazzotta Epifani}, {L{\'o}pez-Moreno},
  {Rodr{\'{\i}}guez}, {Colangeli}, {Palumbo}, {Gr{\"u}n}, {Hilchenbach},
  {Bussoletti}, {Esposito}, {Green}, {Lamy}, {McDonnell}, {Mennella}, {Molina},
  {Morales}, {Moreno}, {Ortiz}, {Palomba}, {Rodrigo}, {Zarnecki}, {Cosi},
  {Giovane}, {Gustafson}, {Herranz}, {Jer{\'o}nimo}, {Leese},
  {L{\'o}pez-Jim{\'e}nez}, \& {Altobelli}}]{2015ApJ...802L..12F}
{Fulle}, M., {Della Corte}, V., {Rotundi}, A., {et~al.} 2015, \apjl, 802, L12

\bibitem[{{Fulle} {et~al.}(2016{\natexlab{a}}){Fulle}, {Della Corte},
  {Rotundi}, {Rietmeijer}, {Green}, {Weissman}, {Accolla}, {Colangeli},
  {Ferrari}, {Ivanovski}, {Lopez-Moreno}, {Epifani}, {Morales}, {Ortiz},
  {Palomba}, {Palumbo}, {Rodriguez}, {Sordini}, \&
  {Zakharov}}]{2016MNRAS.462S.132F}
---. 2016{\natexlab{a}}, \mnras, 462, S132

\bibitem[{{Fulle} {et~al.}(2016{\natexlab{b}}){Fulle}, {Marzari}, {Della
  Corte}, {Fornasier}, {Sierks}, {Rotundi}, {Barbieri}, {Lamy}, {Rodrigo},
  {Koschny}, {Rickman}, {Keller}, {L{\'o}pez-Moreno}, {Accolla}, {Agarwal},
  {A'Hearn}, {Altobelli}, {Barucci}, {Bertaux}, {Bertini}, {Bodewits},
  {Bussoletti}, {Colangeli}, {Cosi}, {Cremonese}, {Crifo}, {Da Deppo},
  {Davidsson}, {Debei}, {De Cecco}, {Esposito}, {Ferrari}, {Giovane},
  {Gustafson}, {Green}, {Groussin}, {Gr{\"u}n}, {Gutierrez}, {G{\"u}ttler},
  {Herranz}, {Hviid}, {Ip}, {Ivanovski}, {Jer{\'o}nimo}, {Jorda},
  {Knollenberg}, {Kramm}, {K{\"u}hrt}, {K{\"u}ppers}, {Lara}, {Lazzarin},
  {Leese}, {L{\'o}pez-Jim{\'e}nez}, {Lucarelli}, {Mazzotta Epifani},
  {McDonnell}, {Mennella}, {Molina}, {Morales}, {Moreno}, {Mottola}, {Naletto},
  {Oklay}, {Ortiz}, {Palomba}, {Palumbo}, {Perrin}, {Rietmeijer},
  {Rodr{\'{\i}}guez}, {Sordini}, {Thomas}, {Tubiana}, {Vincent}, {Weissman},
  {Wenzel}, {Zakharov}, \& {Zarnecki}}]{2016ApJ...821...19F}
{Fulle}, M., {Marzari}, F., {Della Corte}, V., {et~al.} 2016{\natexlab{b}},
  \apj, 821, 19

\bibitem[{{Green} {et~al.}(2004){Green}, {McDonnell}, {McBride}, {Colwell},
  {Tuzzolino}, {Economou}, {Tsou}, {Clark}, \&
  {Brownlee}}]{2004JGRE..10912S04G}
{Green}, S.~F., {McDonnell}, J.~A.~M., {McBride}, N., {et~al.} 2004, Journal of
  Geophysical Research (Planets), 109, E12S04

\bibitem[{{Green} {et~al.}(2007){Green}, {McBride}, {Colwell}, {McDonnell},
  {Tuzzolino}, {Economou}, {Clark}, {Sekanina}, {Tsou}, \&
  {Brownlee}}]{2007ESASP.643...35G}
{Green}, S.~F., {McBride}, N., {Colwell}, M.~T.~S.~H., {et~al.} 2007, Dust in
  Planetary Systems, 643, 35

\bibitem[{{Gr\"un} {et~al.}(1985){Gr\"un}, {Zook}, {Fechtig}, \&
  {Giese}}]{1985Icar...62..244G}
{Gr\"un}, E., {Zook}, H.~A., {Fechtig}, H., \& {Giese}, R.~H. 1985, \icarus,
  62, 244

\bibitem[{{Hanayama} {et~al.}(2012){Hanayama}, {Ishiguro}, {Watanabe},
  {Sarugaku}, {Fukushima}, {Miyaji}, {Yanagisawa}, {Kuroda}, {Yoshida}, {Ohta},
  \& {Nobuyuki}}]{2012PASJ...64..134H}
{Hanayama}, H., {Ishiguro}, M., {Watanabe}, J.-I., {et~al.} 2012, \pasj, 64,
  doi:10.1093/pasj/64.6.134

\bibitem[{{Hilchenbach} {et~al.}(2016){Hilchenbach}, {Kissel}, {Langevin},
  {Briois}, {von Hoerner}, {Koch}, {Schulz}, {Sil{\'e}n}, {Altwegg},
  {Colangeli}, {Cottin}, {Engrand}, {Fischer}, {Glasmachers}, {Gr{\"u}n},
  {Haerendel}, {Henkel}, {H{\"o}fner}, {Hornung}, {Jessberger}, {Lehto},
  {Lehto}, {Raulin}, {Le Roy}, {Ryn{\"o}}, {Steiger}, {Stephan}, {Thirkell},
  {Thomas}, {Torkar}, {Varmuza}, {Wanczek}, {Altobelli}, {Baklouti}, {Bardyn},
  {Fray}, {Kr{\"u}ger}, {Ligier}, {Lin}, {Martin}, {Merouane},
  {Orthous-Daunay}, {Paquette}, {Revillet}, {Siljestr{\"o}m}, {Stenzel}, \&
  {Zaprudin}}]{2016ApJ...816L..32H}
{Hilchenbach}, M., {Kissel}, J., {Langevin}, Y., {et~al.} 2016, \apjl, 816, L32

\bibitem[{{Ipatov} {et~al.}(2008){Ipatov}, {Kutyrev}, {Madsen}, {Mather},
  {Moseley}, \& {Reynolds}}]{2008Icar..194..769I}
{Ipatov}, S.~I., {Kutyrev}, A.~S., {Madsen}, G.~J., {et~al.} 2008, \icarus,
  194, 769

\bibitem[{{Ishiguro} {et~al.}(2007){Ishiguro}, {Sarugaku}, {Ueno}, {Miura},
  {Usui}, {Chun}, \& {Kwon}}]{2007Icar..189..169I}
{Ishiguro}, M., {Sarugaku}, Y., {Ueno}, M., {et~al.} 2007, \icarus, 189, 169

\bibitem[{{Ishiguro} {et~al.}(2013){Ishiguro}, {Yang}, {Usui}, {Pyo}, {Ueno},
  {Ootsubo}, {Minn Kwon}, \& {Mukai}}]{2013ApJ...767...75I}
{Ishiguro}, M., {Yang}, H., {Usui}, F., {et~al.} 2013, \apj, 767, 75

\bibitem[{{Jenniskens} {et~al.}(2016){Jenniskens}, {N{\'e}non}, {Gural},
  {Albers}, {Haberman}, {Johnson}, {Morales}, {Grigsby}, {Samuels}, \&
  {Johannink}}]{2016Icar..266..384J}
{Jenniskens}, P., {N{\'e}non}, Q., {Gural}, P.~S., {et~al.} 2016, \icarus, 266,
  384

\bibitem[{Jeong(2014)}]{2014SNUJeong}
Jeong, J. 2014, Master's thesis, Seoul National University

\bibitem[{{Jewitt} {et~al.}(2015){Jewitt}, {Hsieh}, \&
  {Agarwal}}]{2015aste.book..221J}
{Jewitt}, D., {Hsieh}, H., \& {Agarwal}, J. 2015, {The Active Asteroids}, ed.
  P.~{Michel}, F.~E. {DeMeo}, \& W.~F. {Bottke}, 221--241

\bibitem[{Katyal {et~al.}(2014)Katyal, Banerjee, \& Puri}]{KATYAL2014290}
Katyal, N., Banerjee, V., \& Puri, S. 2014, Journal of Quantitative
  Spectroscopy and Radiative Transfer, 146, 290 , electromagnetic and Light
  Scattering by Nonspherical Particles XIV

\bibitem[{{Kawara} {et~al.}(2017){Kawara}, {Matsuoka}, {Sano}, {Brandt},
  {Sameshima}, {Tsumura}, {Oyabu}, \& {Ienaka}}]{2017PASJ...69...31K}
{Kawara}, K., {Matsuoka}, Y., {Sano}, K., {et~al.} 2017, \pasj, 69, 31

\bibitem[{{Leinert} {et~al.}(1998){Leinert}, {Bowyer}, {Haikala}, {Hanner},
  {Hauser}, {Levasseur-Regourd}, {Mann}, {Mattila}, {Reach}, {Schlosser},
  {Staude}, {Toller}, {Weiland}, {Weinberg}, \& {Witt}}]{1998A&AS..127....1L}
{Leinert}, C., {Bowyer}, S., {Haikala}, L.~K., {et~al.} 1998, \aaps, 127, 1

\bibitem[{{Levison}(1996)}]{1996ASPC..107..173L}
{Levison}, H.~F. 1996, in Astronomical Society of the Pacific Conference
  Series, Vol. 107, Completing the Inventory of the Solar System, ed.
  T.~{Rettig} \& J.~M. {Hahn}, 173--191

\bibitem[{{Levison} \& {Duncan}(1994)}]{1994Icar..108...18L}
{Levison}, H.~F., \& {Duncan}, M.~J. 1994, \icarus, 108, 18

\bibitem[{{Levison} \& {Duncan}(1997)}]{1997Icar..127...13L}
---. 1997, \icarus, 127, 13

\bibitem[{{Mann} {et~al.}(2006){Mann}, {K{\"o}hler}, {Kimura}, {Cechowski}, \&
  {Minato}}]{2006A&ARv..13..159M}
{Mann}, I., {K{\"o}hler}, M., {Kimura}, H., {Cechowski}, A., \& {Minato}, T.
  2006, \aapr, 13, 159

\bibitem[{{Mannel} {et~al.}(2016){Mannel}, {Bentley}, {Schmied}, {Jeszenszky},
  {Levasseur-Regourd}, {Romstedt}, \& {Torkar}}]{2016MNRAS.462S.304M}
{Mannel}, T., {Bentley}, M.~S., {Schmied}, R., {et~al.} 2016, \mnras, 462, S304

\bibitem[{{Mazzotta Epifani} {et~al.}(2009){Mazzotta Epifani}, {Palumbo}, \&
  {Colangeli}}]{2009A&A...508.1031M}
{Mazzotta Epifani}, E., {Palumbo}, P., \& {Colangeli}, L. 2009, \aap, 508, 1031

\bibitem[{{Merouane} {et~al.}(2016){Merouane}, {Zaprudin}, {Stenzel},
  {Langevin}, {Altobelli}, {Della Corte}, {Fischer}, {Fulle}, {Hornung},
  {Sil{\'e}n}, {Ligier}, {Rotundi}, {Ryno}, {Schulz}, {Hilchenbach}, {Kissel},
  \& {Cosima Team}}]{2016A&A...596A..87M}
{Merouane}, S., {Zaprudin}, B., {Stenzel}, O., {et~al.} 2016, \aap, 596, A87

\bibitem[{{Mukai} {et~al.}(1992){Mukai}, {Ishimoto}, {Kozasa}, {Blum}, \&
  {Greenberg}}]{1992A&A...262..315M}
{Mukai}, T., {Ishimoto}, H., {Kozasa}, T., {Blum}, J., \& {Greenberg}, J.~M.
  1992, \aap, 262, 315

\bibitem[{{Nesvorn{\'y}} {et~al.}(2011){Nesvorn{\'y}}, {Janches},
  {Vokrouhlick{\'y}}, {Pokorn{\'y}}, {Bottke}, \&
  {Jenniskens}}]{2011ApJ...743..129N}
{Nesvorn{\'y}}, D., {Janches}, D., {Vokrouhlick{\'y}}, D., {et~al.} 2011, \apj,
  743, 129

\bibitem[{{Nesvorn{\'y}} {et~al.}(2010){Nesvorn{\'y}}, {Jenniskens}, {Levison},
  {Bottke}, {Vokrouhlick{\'y}}, \& {Gounelle}}]{2010ApJ...713..816N}
{Nesvorn{\'y}}, D., {Jenniskens}, P., {Levison}, H.~F., {et~al.} 2010, \apj,
  713, 816

\bibitem[{{Opik}(1951)}]{1951PRIA...54..165O}
{Opik}, E.~J. 1951, Proc.~R.~Irish Acad.~Sect.~A, vol.~54, p.~165-199 (1951).,
  54, 165

\bibitem[{{Pokorn{\'y}} {et~al.}(2014){Pokorn{\'y}}, {Vokrouhlick{\'y}},
  {Nesvorn{\'y}}, {Campbell-Brown}, \& {Brown}}]{2014ApJ...789...25P}
{Pokorn{\'y}}, P., {Vokrouhlick{\'y}}, D., {Nesvorn{\'y}}, D.,
  {Campbell-Brown}, M., \& {Brown}, P. 2014, \apj, 789, 25

\bibitem[{{Rotundi} {et~al.}(2015){Rotundi}, {Sierks}, {Della Corte}, {Fulle},
  {Gutierrez}, {Lara}, {Barbieri}, {Lamy}, {Rodrigo}, {Koschny}, {Rickman},
  {Keller}, {L{\'o}pez-Moreno}, {Accolla}, {Agarwal}, {A'Hearn}, {Altobelli},
  {Angrilli}, {Barucci}, {Bertaux}, {Bertini}, {Bodewits}, {Bussoletti},
  {Colangeli}, {Cosi}, {Cremonese}, {Crifo}, {Da Deppo}, {Davidsson}, {Debei},
  {De Cecco}, {Esposito}, {Ferrari}, {Fornasier}, {Giovane}, {Gustafson},
  {Green}, {Groussin}, {Gr{\"u}n}, {G{\"u}ttler}, {Herranz}, {Hviid}, {Ip},
  {Ivanovski}, {Jer{\'o}nimo}, {Jorda}, {Knollenberg}, {Kramm}, {K{\"u}hrt},
  {K{\"u}ppers}, {Lazzarin}, {Leese}, {L{\'o}pez-Jim{\'e}nez}, {Lucarelli},
  {Lowry}, {Marzari}, {Epifani}, {McDonnell}, {Mennella}, {Michalik}, {Molina},
  {Morales}, {Moreno}, {Mottola}, {Naletto}, {Oklay}, {Ortiz}, {Palomba},
  {Palumbo}, {Perrin}, {Rodr{\'{\i}}guez}, {Sabau}, {Snodgrass}, {Sordini},
  {Thomas}, {Tubiana}, {Vincent}, {Weissman}, {Wenzel}, {Zakharov}, \&
  {Zarnecki}}]{2015Sci...347a3905R}
{Rotundi}, A., {Sierks}, H., {Della Corte}, V., {et~al.} 2015, Science, 347,
  aaa3905

\bibitem[{{Skorov} {et~al.}(2016){Skorov}, {Reshetnyk}, {Lacerda}, {Hartogh},
  \& {Blum}}]{2016MNRAS.461.3410S}
{Skorov}, Y., {Reshetnyk}, V., {Lacerda}, P., {Hartogh}, P., \& {Blum}, J.
  2016, \mnras, 461, 3410

\bibitem[{{Snodgrass} {et~al.}(2011){Snodgrass}, {Fitzsimmons}, {Lowry}, \&
  {Weissman}}]{2011MNRAS.414..458S}
{Snodgrass}, C., {Fitzsimmons}, A., {Lowry}, S.~C., \& {Weissman}, P. 2011,
  \mnras, 414, 458

\bibitem[{{Soja} {et~al.}(2016){Soja}, {Schwarzkopf}, {Sommer}, {Vaubaillon},
  {Albin}, {Rodmann}, {Gr{\"u}n}, \& {Srama}}]{2016pimo.conf..284S}
{Soja}, R.~H., {Schwarzkopf}, G.~J., {Sommer}, M., {et~al.} 2016, in
  International Meteor Conference Egmond, the Netherlands, 2-5 June 2016, ed.
  A.~{Roggemans} \& P.~{Roggemans}, 284--286

\bibitem[{{Steel} \& {Elford}(1986)}]{1986MNRAS.218..185S}
{Steel}, D.~I., \& {Elford}, W.~G. 1986, \mnras, 218, 185

\bibitem[{{Tancredi} {et~al.}(2006){Tancredi}, {Fern{\'a}ndez}, {Rickman}, \&
  {Licandro}}]{2006Icar..182..527T}
{Tancredi}, G., {Fern{\'a}ndez}, J.~A., {Rickman}, H., \& {Licandro}, J. 2006,
  \icarus, 182, 527

\bibitem[{{Ueda} {et~al.}(2017){Ueda}, {Kobayashi}, {Takeuchi}, {Ishihara},
  {Kondo}, \& {Kaneda}}]{2017AJ....153..232U}
{Ueda}, T., {Kobayashi}, H., {Takeuchi}, T., {et~al.} 2017, \aj, 153, 232

\bibitem[{{Vaubaillon} \& {Reach}(2010)}]{2010AJ....139.1491V}
{Vaubaillon}, J.~J., \& {Reach}, W.~T. 2010, \aj, 139, 1491

\bibitem[{{Wiegert} {et~al.}(2009){Wiegert}, {Vaubaillon}, \&
  {Campbell-Brown}}]{2009Icar..201..295W}
{Wiegert}, P., {Vaubaillon}, J., \& {Campbell-Brown}, M. 2009, \icarus, 201,
  295

\bibitem[{{Yang} \& {Ishiguro}(2015)}]{2015ApJ...813...87Y}
{Yang}, H., \& {Ishiguro}, M. 2015, \apj, 813, 87

\end{thebibliography}

\end{document}